\documentclass[12pt,letter]{article}
\usepackage[cmex10]{amsmath}
\usepackage{amsfonts}
\usepackage{IEEEtrantools}
\usepackage{diagrams}
\usepackage{array}
\title{The Dirac equation in curved spacetimes using coordinate-free notation}
\author{Mihai Moise \\
\\
email: mihai.moise5@gmail.com}
\usepackage[bookmarks=false]{hyperref}
\begin{document}
\maketitle
\begin{abstract}
\noindent The Dirac equation in curved spacetimes is formulated using co\-or\-di\-nate-free notation. A Lagrangean density which corresponds to the subject equation is presented. The subject equation is invariant under a local rotation of the coframe. The current is independent of the local orientation of the coframe and it is conserved. The subject equation has an equivalent formulation which uses the Christoffel gamma. The stress-energy tensor is calculated.
\end{abstract}

\section{Introduction}

The Dirac equation in curved spacetimes has previously been studied \cite{Visinescu} \cite{Weyl} \cite{Marchuk}. A coordinate-free form different from the one in this article is in \cite{Marchuk2}. In this article, the Dirac equation in curved spacetimes is formulated using coordinate-free notation. The layout of this article is: In section \ref{Lagrangean density}, I show that the Dirac equation in curved spacetimes is derived from the Dirac Lagrangean. In section \ref{Coframe independance}, I show that the subject equation is invariant under a local rotation of the coframe. In section \ref{Conservation of current}, I show that the current is conserved. In section \ref{Alternative formulation}, I show that the subject equation has an equivalent formulation which uses the Christoffel gamma. In section \ref{Stress-energy tensor}, the stress-energy tensor is calculated. The \hyperref[Appendix]{Appendix} contains a proof that $\omega$ has the same values as the rotational part of the torsion-free Cartan connection $(\theta, \omega): \mathrm{T}(\mathcal{M}^4) \rightarrow \mathfrak{euc}(3,1)$. 

For a curved spacetimes Dirac equation, we usually use a coframe 1-form $\theta: \mathrm{T}(\mathcal{M}^4) \rightarrow \mathbb{R}^4$ and a dual vector field $\theta^j(v_k) = \delta^j_{\phantom{j}k}$. I use,
\begin{equation}
\begin{array}{ccc}
\sigma^1 = \left[\begin{array}{cc} 0 & 1 \\ 1 & 0 \end{array} \right] &
\sigma^2 = \left[\begin{array}{cc} 0 & -i \\ i & 0 \end{array} \right] &
\sigma^3 = \left[\begin{array}{cc} 1 & 0 \\ 0 & -1 \end{array} \right]
\end{array}
\end{equation}
\begin{equation}
\begin{array}{cc}
\gamma^0 = \left[\begin{array}{cc} i & 0 \\ 0 & -i \end{array} \right] &
\gamma^j = \left[\begin{array}{cc} 0 & \sigma^j \\ \sigma^j & 0 \end{array} \right]
\end{array}
\end{equation}
\begin{IEEEeqnarray}{rCl}
\eta & = & \left[\begin{array}{cccc} -1 & 0 & 0 & 0 \\
0 & 1 & 0 & 0 \\
0 & 0 & 1 & 0 \\
0 & 0 & 0 & 1 \end{array} \right] \\
\gamma^a \gamma^b + \gamma^b \gamma^a & = & 2 \eta^{ab} \\
\overline{\Psi} & = & \Psi^\dagger \gamma^0 \\
S^{ab} & = & \frac{1}{4} [\gamma^a, \gamma^b] \\
\epsilon^{0123} & = & \epsilon_{0123}
\end{IEEEeqnarray}
Latin indices are Minkowskian and Greek indices are spacetime indices. Latin indices are raised or lowered using the Minkowski metric:
\begin{equation}
A^a = \eta^{ab} A_b
\end{equation}

The coframe $\theta$ can be expressed using the coordinate 1-forms,
\begin{equation}
\theta^a = \theta^a_{\phantom{a}\mu} \mathrm{d}x^\mu
\end{equation}
and the metric is,
\begin{equation}
g_{\mu\nu} = \eta_{ab} \theta^a_{\phantom{a}\mu} \theta^b_{\phantom{b}\nu} \label{eq10}
\end{equation}
\eqref{eq10} is 10 equations for 16 components, so for a given metric, $\theta$ has 6 degrees of freedom.

The Dirac equation in curved spacetimes is,
\begin{gather}
\left(\gamma^a v_a + ie\gamma^a A (v_a) - \frac{1}{4} \gamma^a \gamma^b \gamma^c \omega_{bc}(v_a) + m \right) \Psi = 0 \\
\omega_{bc}(v_a) = \frac{1}{2} \Big( \mathrm{d}\theta_a(v_b, v_c) + \mathrm{d}\theta_b(v_a, v_c) - \mathrm{d}\theta_c(v_a, v_b) \Big)
\end{gather}
In differential geometry style \cite{Sharpe}, vectors are represented as derivatives, so $v_a \Psi$ is an equivalent notation to $\mathrm{d} \Psi (v_a)$. $\omega$ is antisymmetric in its two indices:
\begin{equation}
\omega_{bc} = - \omega_{cb}
\end{equation}
so these two terms are equal:
\begin{equation}
- \frac{1}{4} \gamma^a \gamma^b \gamma^c \omega_{bc}(v_a) = - \frac{1}{2} \gamma^a S^{bc} \omega_{bc}(v_a)
\end{equation}
$e$ is the electric charge of the particle. $A$ is the electromagnetic potential represented as a 1-form. An alternative formulation for the Dirac equation in curved spacetimes is,
\begin{equation}
\left(\gamma^a v_a^{\phantom{a}\mu} \partial_\mu + ie\gamma^a A (v_a) - \frac{1}{2} \gamma^a S^{bc} v_a^{\phantom{a} \mu} v_b^{\phantom{b} \nu}\left( - \partial_\mu\theta_{c \nu} + \theta_{c  \rho} \Gamma^\rho_{\mu \nu}\right) + m \right) \Psi = 0 
\end{equation}
 
As a side note, $\omega$ has the same values as the rotational part of the torsion-free Cartan connection $(\theta, \omega): \mathrm{T}(\mathcal{M}^4) \rightarrow \mathfrak{euc}(3,1)$ because $\omega$ is antisymmetric and follows the equation,
\begin{equation}
\mathrm{d}\theta^a - \omega^a_{\phantom{a}b} \wedge \theta^b = 0 
\end{equation}
The proof is in the \hyperref[Appendix]{Appendix}.

In this article, the Hodge dual is defined as,
\begin{equation}
* \left( \theta_{i_0} \wedge \ldots \wedge \theta_{i_{k - 1}} \right) = 
\frac{1}{(n-k)!}\epsilon_{i_0 .. i_{k-1} j_0 .. j_{n - k - 1}} \theta^{j_0} \wedge \ldots \wedge \theta^{j_{n - k - 1}}
\end{equation}
This means that,
\begin{IEEEeqnarray}{rCl}
* 1 & = & \theta^0 \wedge \theta^1 \wedge \theta^2 \wedge \theta^3 \\
*\!* \! 1 & = & * \left( \theta^0 \wedge \theta^1 \wedge \theta^2 \wedge \theta^3 \right) \\ 
& = & * \left( \eta^{00} \eta^{11} \eta^{22} \eta^{33} \theta_0 \wedge \theta_1 \wedge \theta_2 \wedge \theta_3 \right)\\
& = & - *  \left( \theta_0 \wedge \theta_1 \wedge \theta_2 \wedge \theta_3 \right)\\
& = & -1 
\end{IEEEeqnarray}
The current is usually defined as,
\begin{equation}
j = - e\overline{\Psi} \gamma^a \Psi * \! \theta_a
\end{equation}
In curved spacetimes, we would like the current to be conserved,
\begin{equation}
\mathrm{d} j = 0
\end{equation}
We would also like that if $\Psi_1$ is a solution of the curved spacetimes Dirac equation with coframe $\theta_1$, and $\theta_2$ another coframe for the same metric, then for a solution $\Psi_2$ of the curved spacetimes Dirac equation with coframe $\theta_2$ to exist that has the same current:
\begin{IEEEeqnarray}{rCl}
j_1 & = & - e\overline{\Psi_1} \gamma^a \Psi_1 * \! \theta_{1a} \\
j_2 & = & - e\overline{\Psi_2} \gamma^a \Psi_2 * \! \theta_{2a} \\
j_1 & = & j_2
\end{IEEEeqnarray}

\section{Lagrangean density} \label{Lagrangean density}

The Lagrangean density which corresponds to the Dirac equation in curved spacetimes is,
\begin{IEEEeqnarray}{c}
\IEEEeqnarraymulticol{1}{l}{\mathfrak{L} = -\frac{1}{2} \mathrm{d} A \wedge * \mathrm{d} A\:{+} } \notag \\
\IEEEeqnarraymulticol{1}{r} { {+}\:i \overline{\Psi} \left(\gamma^a \mathrm{d}\Psi \wedge * \theta_a + ie\overline{\Psi} \gamma^a A \wedge *\theta_a \Psi- \frac{1}{4}\gamma^a \gamma^b \gamma^c  \omega_{bc} \wedge *\theta_a \Psi + m \Psi * \! 1 \right) } \IEEEeqnarraynumspace
\end{IEEEeqnarray}
To show that this Lagrangean density corresponds to the Dirac equation in curved spacetimes, I proceed in the usual manner, setting up a region of spacetime and a small variation of the field $\delta \Psi$ which vanishes on the region boundary. The variation of the Lagrangean density is,
\begin{IEEEeqnarray}{c}
\IEEEeqnarraymulticol{1}{l}{\delta \mathfrak{L} = i \delta \overline{\Psi} \left(\gamma^a \mathrm{d}\Psi \wedge * \theta_a + ie\gamma^a A \wedge *\theta_a \Psi - \frac{1}{4}\gamma^a \gamma^b \gamma^c  \omega_{bc} \wedge *\theta_a \Psi + m \Psi * \! 1 \right)\:{+} } \notag \\
\IEEEeqnarraymulticol{1}{r}{{+}\:i\overline{\Psi} \left(\gamma^a \mathrm{d} \delta \Psi \wedge * \theta_a + ie\gamma^a A \wedge *\theta_a \delta \Psi- \frac{1}{4}\gamma^a \gamma^b \gamma^c  \omega_{bc} \wedge *\theta_a \delta\Psi + m * \! 1 \delta \Psi\right) } \notag \\
\end{IEEEeqnarray}
I have,
\begin{equation}
\mathrm{d} \left(\overline{\Psi}\gamma^a\delta \Psi * \! \theta_a \right) = \mathrm{d}\overline{\Psi} \wedge * \theta_a \gamma^a \delta \Psi + \overline{\Psi} \gamma^a \mathrm{d} \delta \Psi \wedge * \theta_a + \overline{\Psi} \gamma^a  \delta \Psi \, \mathrm{d} \! * \! \theta_a
\end{equation}
so,
\begin{equation}
\overline{\Psi} \gamma^a \mathrm{d} \delta \Psi \wedge * \theta_a = \mathrm{d} \left(\overline{\Psi}\gamma^a\delta \Psi * \! \theta_a \right) - \mathrm{d}\overline{\Psi} \wedge * \theta_a \gamma^a \delta \Psi - \overline{\Psi} \gamma^a  \delta \Psi \, \mathrm{d} \! * \! \theta_a
\end{equation}
Substituting this, I get,
\begin{multline}
\delta \mathfrak{L}  =  i\delta\overline{\Psi}\left(\gamma^a \mathrm{d}\Psi \wedge * \theta_a + ie\gamma^a A \wedge *\theta_a \Psi - \frac{1}{4}\gamma^a \gamma^b \gamma^c  \omega_{bc} \wedge *\theta_a \Psi + m \Psi * \! 1 \right)\:{+} \\
{+}\:i\Big( -\mathrm{d}\overline{\Psi} \wedge * \theta_a \gamma^a  - \overline{\Psi} \gamma^a  \mathrm{d} \! * \! \theta_a + ie\overline{\Psi}\gamma^a A \wedge *\theta_a - \frac{1}{4}\overline{\Psi} \gamma^a \gamma^b \gamma^c  \omega_{bc} \wedge *\theta_a\:{+} \\
{+}\:m \overline{\Psi}* \! 1 \Big) \delta \Psi + \mathrm{d} \left(i\overline{\Psi}\gamma^a\delta \Psi * \! \theta_a \right) 
\end{multline}
Using \eqref{eq104} - \eqref{eq153b}, it can be gathered that,
\begin{IEEEeqnarray}{rCl}
\gamma^a\gamma^b\gamma^c \omega_{bc}(v_a) * \! 1 & = & - \gamma^c\gamma^b\gamma^a \omega_{bc}(v_a) * \! 1 - 4 \gamma^b \mathrm{d}\theta^a(v_b, v_a) * \! 1 \\
\mathrm{d} \! * \! \theta_0 & = & \mathrm{d} ( \theta^1 \wedge \theta^2 \wedge \theta^3 ) \\
& = & \mathrm{d} \theta^1 \wedge \theta^2 \wedge \theta^3 - \theta^1 \wedge \mathrm{d} \theta^2 \wedge \theta^3 + \ldots \\
& = & \mathrm{d} \theta^1(v_0, v_1) * \! 1 + \mathrm{d} \theta^2(v_0, v_2) * \! 1 + \ldots \\
\mathrm{d} \! * \! \theta_a & = & \mathrm{d} \theta^b (v_a, v_b) * \! 1 \\
\gamma^a\gamma^b\gamma^c \omega_{bc} \wedge *\theta_a & = & - \gamma^c\gamma^b\gamma^a \omega_{bc}\wedge *\theta_a - 4 \gamma^a \mathrm{d} \! * \! \theta_a
\end{IEEEeqnarray}
Substituting this, I get,
\begin{multline}
\!\!\!\!\delta \mathfrak{L} = i \delta \overline{\Psi} \left(\gamma^a \mathrm{d}\Psi \wedge * \theta_a + ie\gamma^a A \wedge *\theta_a \Psi - \frac{1}{4}\gamma^a \gamma^b \gamma^c  \omega_{bc} \wedge *\theta_a \Psi + m \Psi * \! 1 \right)\:{+} \\
{+}\:i\left( -\mathrm{d}\overline{\Psi} \wedge * \theta_a \gamma^a + ie\overline{\Psi}\gamma^a A \wedge *\theta_a + \frac{1}{4}\overline{\Psi} \gamma^c \gamma^b \gamma^a  \omega_{bc} \wedge *\theta_a
+  m \overline{\Psi}* \! 1 \right) \delta \Psi\:{+}  \\
{+}\:\mathrm{d} \left(i\overline{\Psi}\gamma^a\delta \Psi * \! \theta_a \right) 
\end{multline}
Comparing with the Dirac equation in curved spacetimes and its Hermitean conjugate \eqref{eq168},
\begin{multline}
\delta \mathfrak{L} = 2 \mathrm{Re} \bigg(i \delta \overline{\Psi} \Big(\gamma^a \mathrm{d}\Psi \wedge * \theta_a + ie\gamma^a A \wedge *\theta_a \Psi - \frac{1}{4}\gamma^a \gamma^b \gamma^c  \omega_{bc} \wedge *\theta_a \Psi\:{+} \\
{+}\:m \Psi * \! 1 \Big)\bigg) + 
\mathrm{d} \left(i\overline{\Psi}\gamma^a\delta \Psi * \! \theta_a \right) 
\end{multline}
The variation of the Lagrangean density must vanish up to a 4-divergence for any small value of $\delta \overline{\Psi}$, so this Lagrangean density implies the Dirac equation in curved spacetimes. Because the variation of the Lagrangean density is real up to a 4-divergence, the following Lagrangean density implies the same field equation:
\begin{IEEEeqnarray}{c}
\IEEEeqnarraymulticol{1}{l}{ \mathfrak{L} = -\frac{1}{2} \mathrm{d} A \wedge * \mathrm{d} A\:{+} } \notag \\
\IEEEeqnarraymulticol{1}{r}{ {+}\:\mathrm{Re} \bigg(i \overline{\Psi} \left(\gamma^a \mathrm{d}\Psi \wedge * \theta_a + ie\overline{\Psi} \gamma^a A \wedge *\theta_a \Psi- \frac{1}{4}\gamma^a \gamma^b \gamma^c  \omega_{bc} \wedge *\theta_a \Psi + m \Psi * \! 1 \right) \bigg) } \notag \\
\end{IEEEeqnarray}
Using,
\begin{IEEEeqnarray}{rCl}
\mathrm{Re}(z) & = & \: \frac{z + z^{\dagger}}{2} \\
D_{\mu} & = & \: \partial_{\mu} + ieA_{\mu} - \frac{1}{2} S^{bc} \omega_{bc}(\partial_{\mu}) \\
\gamma^{a \dagger} & = & \left\{ \begin{IEEEeqnarraybox}[][c]{l?l}
\IEEEstrut
- \gamma^a & \text{if } a = 0 \\
\gamma^a & \text{if } a \neq 0
\IEEEstrut
\end{IEEEeqnarraybox} \right. \\
\gamma^{a \dagger} \gamma^0 & = & - \gamma^0 \gamma^a \\
\left(\frac{i}{2} \overline{\Psi} \gamma^a v_{a}^{\phantom{a}{\mu}} D_{\mu} \Psi \right)^{\dagger} & = & \: \frac{-i}{2} (D_{\mu} \Psi)^{\dagger} \gamma^{a\dagger} v_{a}^{\phantom{a}{\mu}} \gamma^{0\dagger} \Psi \\
& = & \: \frac{-i}{2} (D_{\mu} \Psi)^{\dagger} \gamma^{a\dagger} v_{a}^{\phantom{a}{\mu}} (-\gamma^0) \Psi \\
& =  & \: \frac{-i}{2} (D_{\mu} \Psi)^{\dagger}  \gamma^0 \gamma^a v_{a}^{\phantom{a}{\mu}} \Psi \\
& =  & \: \frac{-i}{2} (\overline{D_{\mu} \Psi}) \gamma^a v_{a}^{\phantom{a}{\mu}} \Psi
\end{IEEEeqnarray}
The previous Lagrangean density corresponds to the action,
\begin{equation}
S = \int \mathrm{d}x^n \sqrt{g} \left(-\frac{1}{4} F^{\mu\nu} F_{\mu\nu} + \frac{i}{2} \Big(\overline{\Psi} \gamma^a v_{a}^{\phantom{a}{\mu}} D_{\mu} \Psi
- (\overline{D_{\mu} \Psi}) \gamma^a v_{a}^{\phantom{a}{\mu}} \Psi \Big) + im \overline{\Psi} \Psi \right) 
\end{equation}
with $g = | \mathrm{det}(g_{\mu\nu}) |$.

\section{Coframe independance} \label{Coframe independance}

I shall show that the Dirac equation in curved spacetimes is invariant under a  local rotation of the coframe. The coframes $\theta_2$ and $\theta_1$ are related by,
\begin{equation}
\theta_{2}^{\phantom{2}a} = \Lambda(x)^a_{\phantom{a}b} \theta_{1}^{\phantom{1}b}
\end{equation}
With $\Lambda \in \mathrm{SO}(3,1)$.  $\Lambda$ can have different values at different events in spacetime. The vector fields $v_2$ and $v_1$ are related by,
\begin{equation}
v_{2a} = \Lambda_a^{\phantom{a}b} v_{1b}
\end{equation}
That this works can be checked:
\begin{IEEEeqnarray}{rCl}
\theta_{2}^{\phantom{2}a}(v_{2c}) & = & \Lambda^a_{\phantom{a}b} \theta_{1}^{\phantom{2}b} ( \Lambda_c^{\phantom{c}e} v_{1e} ) \\
& = & \Lambda^a_{\phantom{a}b} \Lambda_c^{\phantom{c}e} \delta^b_{\phantom{b}e} \\
& = & \Lambda^a_{\phantom{a}b} \Lambda_c^{\phantom{c}b} \\
& = & \delta^a_{\phantom{a}c}
\end{IEEEeqnarray}
The last equation holds because $\Lambda$ is in $\mathrm{SO}(3,1)$.
$\Lambda$ can be expressed as,
\begin{equation}
\Lambda^a_{\phantom{a}b} = \big(\mathrm{exp}(k^c_{\phantom{c}d}\mathfrak{I}_c^{\phantom{c}d}) \big)^a_{\phantom{a}b}
\end{equation}
I use the symbol $\mathfrak{I}_c^{\phantom{c}d}$ to mean a matrix with a $1$ at row $c$ column $d$ and $0$ at every other entry. The notation $()^a_{\phantom{a}b}$ means row $a$ column $b$ of the matrix inside the parentheses. $k$ is antisymmetric:
\begin{equation}
k_{ab} = - k_{ba}
\end{equation}

I suppose that there is a solution $\Psi_1$ with coframe $\theta_1$ and look for another solution $\Psi_2$ with coframe $\theta_2$ which has the same current:
\begin{IEEEeqnarray}{rCl}
j_1 & = & j_2 \\
- e\overline{\Psi_1} \gamma^a \Psi_1 v_{1a} & = & - e\overline{\Psi_2} \gamma^a \Psi_2 v_{2a} \\
- e\overline{\Psi_1} \gamma^a \Psi_1 v_{1a} & = &
- e\overline{\Psi_2} \gamma^a \Psi_2 \Lambda_a^{\phantom{a}b} v_{1b}
\end{IEEEeqnarray}
A solution is,
\begin{equation}
\Psi_2 = \Lambda_{\frac{1}{2}}(x) \Psi_1
\end{equation}
with,
\begin{equation}
\Lambda_{\frac{1}{2}}^{\phantom{\frac{1}{2}}-1} \gamma^a \Lambda_{\frac{1}{2}} = \Lambda^a_{\phantom{a}b} \gamma^b \label{eq48}
\end{equation}
$\Lambda_{\frac{1}{2}}$ can have different values at different events in spacetime. Later in this article, I shall give a method of constructing $\Lambda_{\frac{1}{2}}$. The equation becomes,
\begin{equation}
- \overline{\Psi_1} \gamma^a \Psi_1 v_{1a} =
- \Psi_1^\dagger \Lambda_{\frac{1}{2}}^\dagger \gamma^0 \gamma^a \Lambda_{\frac{1}{2}} \Psi_1 \Lambda_a^{\phantom{a}b} v_{1b}
\end{equation}
I have,
\begin{IEEEeqnarray}{rCl}
\Lambda_{\frac{1}{2}}^{\phantom{\frac{1}{2}}\dagger} \gamma^0 & = & \mathrm{exp}\left(\frac{1}{8}k_{ab} [\gamma^a, \gamma^b] \right)^\dagger \gamma^0\\
& = & \left(1 + \frac{1}{8}k_{ab}[\gamma^a, \gamma^b] + \frac{1}{128}k_{ab}[\gamma^a, \gamma^b]k_{cd}[\gamma^c, \gamma^d] + \ldots\right)^\dagger \gamma^0\\
& = & \left(1 + \frac{1}{8}k_{ab}[\gamma^{b\dagger}, \gamma^{a\dagger}] + \frac{1}{128}k_{cd}[\gamma^{d\dagger}, \gamma^{c\dagger}]k_{ab}[\gamma^{b\dagger}, \gamma^{a\dagger}] + \ldots\right)\gamma^0 \IEEEeqnarraynumspace \\
& = & \gamma^0\left(1 + \frac{1}{8}k_{ab}[\gamma^{b}, \gamma^{a}] + \frac{1}{128}k_{cd}[\gamma^{d}, \gamma^{c}]k_{ab}[\gamma^{b}, \gamma^{a}] + \ldots\right) \\
& = & \gamma^0\left(1 - \frac{1}{8}k_{ab}[\gamma^{a}, \gamma^{b}] + \frac{1}{128}k_{ab}[\gamma^{a}, \gamma^{b}]k_{cd}[\gamma^{c}, \gamma^{d}] + \ldots\right) \\
& = & \gamma^0 \Lambda_{\frac{1}{2}}^{\phantom{\frac{1}{2}}-1}
\end{IEEEeqnarray}
then the equation becomes,
\begin{IEEEeqnarray}{rCl}
- \overline{\Psi_1} \gamma^a \Psi_1 v_{1a} & = &
- \Psi_1^\dagger \gamma^0 \Lambda_{\frac{1}{2}}^{\phantom{\frac{1}{2}}-1}  \gamma^a \Lambda_{\frac{1}{2}} \Psi_1 \Lambda_a^{\phantom{a}b} v_{1b} \\
& = & - \overline{\Psi_1}   \Lambda^a_{\phantom{a}c} \gamma^c  \Psi_1 \Lambda_a^{\phantom{a}b} v_{1b} \\
& = & - \overline{\Psi_1} \gamma^a  \Psi_1 v_{1a}
\end{IEEEeqnarray}
The last equation holds because $\Lambda$ is in $\mathrm{SO}(3,1)$.

Coframe transformations can be composited. Given two coframe transformations $(\Lambda_1, \Lambda_{\frac{1}{2}1}): (\theta_1, \Psi_1) \rightarrow (\theta_2, \Psi_2)$ and $(\Lambda_2, \Lambda_{\frac{1}{2}2}): (\theta_2, \Psi_2) \rightarrow (\theta_3, \Psi_3)$, there is a composite coframe transformation $(\Lambda_3, \Lambda_{\frac{1}{2}3}): (\theta_1, \Psi_1) \rightarrow (\theta_3, \Psi_3)$ which makes this diagram commute:
\begin{diagram}
(\theta_1, \Psi_1) & \rTo^{(\Lambda_1, \Lambda_{\frac{1}{2}1})} & (\theta_2, \Psi_2) \\
& \rdTo^{(\Lambda_3, \Lambda_{\frac{1}{2}3})} & \dTo_{(\Lambda_2, \Lambda_{\frac{1}{2}2})} \\
&  & (\theta_3, \Psi_3)
\end{diagram}
It is,
\begin{IEEEeqnarray}{rCl}
\Lambda_{3\phantom{a}b}^{\phantom{3}a} & = & \Lambda_{2\phantom{a}c}^{\phantom{1}a} \Lambda_{1\phantom{c}b}^{\phantom{2}c} \\
\Lambda_{\frac{1}{2}3} & = & \Lambda_{\frac{1}{2}2} \Lambda_{\frac{1}{2}1}
\end{IEEEeqnarray}
Because the end values of the fields $\theta$ and $\Psi$ are the same for whichever path is taken in the previous commutative diagram, and $\theta$ and $\Psi$ are the only fields which enter the current and the Lagrangean which are affected by a coframe transformation, it follows that $(\Lambda_3, \Lambda_{\frac{1}{2}3})$ leave the current and the Lagrangean invariant if  $(\Lambda_1, \Lambda_{\frac{1}{2}1})$ and $(\Lambda_2, \Lambda_{\frac{1}{2}2})$ leave the current and the Lagrangean invariant.

An arbitrary coframe transformation $\Lambda_{\mathrm{arbitrary}}$ can be decomposed into Euler angles:
\begin{equation}
\Lambda_{\mathrm{arbitrary}} = \Lambda_6 \Lambda_5 \Lambda_4 \Lambda_3 \Lambda_2 \Lambda_1 \label{eq73}
\end{equation}
The coframe transformations $\Lambda_1$ through $\Lambda_6$ have all but at most one of the $k_{ab}$ equal to 0 everywhere in spacetime. Without loss of generality, I can assume that for $\Lambda$ all but at most one of the $k_{ab}$ are zero everywhere in spacetime. $\Lambda_{\frac{1}{2}}$ defined as,
\begin{equation}
\Lambda_{\frac{1}{2}} = \mathrm{exp}\left(\frac{1}{2}k_{ab} S^{ab} \right) \label{eq74}
\end{equation}
follows \eqref{eq48}. To show it, I denote $r$ and $s$ the fixed indices of the Euler angle with $r < s$. If $k_{rs}$ is a spatial rotation, then,
\begin{IEEEeqnarray}{rCl}
\mathrm{exp}\left(\frac{1}{2}k_{ab} S^{ab} \right) & = & \mathrm{exp}\left(\frac{1}{2} k_{rs} \gamma^r \gamma^s \right) \\
& = &  \cos \left(\frac{k_{rs}}{2}\right) + \sin \left( \frac{k_{rs}}{2}\right) \gamma^r \gamma^s
\end{IEEEeqnarray}
\begin{IEEEeqnarray}{rCl}
\IEEEeqnarraymulticol{3}{l}{\Lambda_{\frac{1}{2}}^{\phantom{\frac{1}{2}}-1} \gamma^a \Lambda_{\frac{1}{2}} = } \notag \\ 
 \quad & = &  \left(\cos \left(\frac{k_{rs}}{2} \right) - \sin \left( \frac{k_{rs}}{2} \right) \gamma^r \gamma^s \right)    \gamma^a\:{\times} \notag \\
 & & {\times}\:\left(\cos \left(\frac{k_{rs}}{2} \right) + \sin \left( \frac{k_{rs}}{2} \right) \gamma^r \gamma^s \right) \\
& = & \cos \left(\frac{k_{rs}}{2} \right)^2 \gamma^a + \cos \left(\frac{k_{rs}}{2} \right) \sin \left( \frac{k_{rs}}{2} \right)  \big(\gamma^a \gamma^r \gamma^s - \gamma^r \gamma^s \gamma^a \big)\:{-} \\
& &  {-}\:\sin \left( \frac{k_{rs}}{2} \right)^2 \gamma^r \gamma^s  \gamma^a \gamma^r \gamma^s \\
& = & \cos \left(\frac{k_{rs}}{2} \right)^2 \gamma^a + \cos \left(\frac{k_{rs}}{2} \right)\sin \left( \frac{k_{rs}}{2} \right) \big( 2\eta^{ar} \gamma^s - 2\eta^{as}\gamma^r \big)\:{-} \\
& & {-}\:\sin \left( \frac{k_{rs}}{2} \right)^2 \big(\gamma^r \eta^{ar} \eta^{ss} + \gamma^s \eta^{as} \eta^{rr} - \gamma^a \eta^{rr} \eta^{ss} (1-\delta^a_{\phantom{a}r})(1-\delta^a_{\phantom{a}s})\big) \IEEEeqnarraynumspace \\
\noalign{\noindent If $a$ = $r$, \vspace*{\jot}}
& = &  \bigg(\cos \left(\frac{k_{rs}}{2} \right)^2 - \sin \left( \frac{k_{rs}}{2} \right)^2 \bigg) \gamma^a + 2 \cos \left(\frac{k_{rs}}{2} \right) \sin \left( \frac{k_{rs}}{2} \right) \gamma^s \\
& = & \cos(k_{rs})\gamma^a + \sin( k_{rs}) \gamma^s
\end{IEEEeqnarray}
\begin{IEEEeqnarray}{rCl}
\Lambda^a_{\phantom{a}b} \gamma^b & = & \Big(\mathrm{exp}\big(k_{rs}(\mathfrak{I}^{rs} - \mathfrak{I}^{sr})\big) \Big)^a_{\phantom{a}b} \gamma^b\\
& = & \Big(\big(\cos(k_{rs}) - 1\big)\big(\mathfrak{I}_r^{\phantom{r}r} + \mathfrak{I}_s^{\phantom{s}s}\big) + \delta^c_{\phantom{c}d}\mathfrak{I}_c^{\phantom{c}d} + \sin( k_{rs})\big(\mathfrak{I}^{rs} - \mathfrak{I}^{sr}\big)\Big)^a_{\phantom{a}b} \gamma^b \IEEEeqnarraynumspace \\  
\noalign{\noindent If $a = r$, \vspace{\jot}}
& = & \cos(k_{rs})\gamma^a + \sin( k_{rs}) \gamma^s 
\end{IEEEeqnarray}
The case $a = s$ is similar by antisymmetry. If $k_{rs}$ is a boost, the proof is similar, \textit{mutatis mutandis}.

If $\Lambda_{1}$, $\Lambda_{2}$, $\Lambda_{3}$, $\Lambda_{\frac{1}{2}1}$,  $\Lambda_{\frac{1}{2}2}$, $\Lambda_{\frac{1}{2}3}$, are such that,
\begin{IEEEeqnarray}{rCl}
\Lambda_{\frac{1}{2}1}^{\phantom{\frac{1}{2}1}-1} \gamma^a \Lambda_{\frac{1}{2}1} & = & \Lambda^{\phantom{1}a}_{1\phantom{a}b} \gamma^b \\
\Lambda_{\frac{1}{2}2}^{\phantom{\frac{1}{2}2}-1} \gamma^a \Lambda_{\frac{1}{2}2} & = & \Lambda^{\phantom{2}a}_{2\phantom{a}b} \gamma^b \\
\Lambda_{3\phantom{a}b}^{\phantom{3}a} & = &  \Lambda_{2\phantom{a}c}^{\phantom{1}a} \Lambda_{1\phantom{c}b}^{\phantom{2}c}  \\
\Lambda_{\frac{1}{2}3} & = & \Lambda_{\frac{1}{2}2} \Lambda_{\frac{1}{2}1} \\
\noalign{\noindent then, \vspace*{\jot}}
\Lambda_{\frac{1}{2}3}^{\phantom{\frac{1}{2}3}-1} \gamma^a \Lambda_{\frac{1}{2}3} & = & \Lambda_{\frac{1}{2}2}^{\phantom{\frac{1}{2}2}-1} \Lambda_{\frac{1}{2}2}^{\phantom{\frac{1}{2}2}-1} \gamma^a \Lambda_{\frac{1}{2}2} \Lambda_{\frac{1}{2}1} \\
 & = & \Lambda_{\frac{1}{2}2}^{\phantom{\frac{1}{2}2}-1}\Lambda^{\phantom{2}a}_{2\phantom{a}b} \gamma^b \Lambda_{\frac{1}{2}1} \\
 & = & \Lambda^{\phantom{2}a}_{2\phantom{a}b} \Lambda^{\phantom{1}b}_{1\phantom{b}c} \gamma^c \\
 & = & \Lambda^{\phantom{3}a}_{3\phantom{a}c} \gamma^c 
\end{IEEEeqnarray}

An algorithm for assigning values to $\Lambda_{\frac{1}{2}\mathrm{arbitrary}}$ is to decompose $\Lambda_{\mathrm{arbitrary}}$ into Euler angles as in \eqref{eq73}, then assigning values to $\Lambda_{\frac{1}{2}1}$ through $\Lambda_{\frac{1}{2}6}$ using \eqref{eq74}, then calculating, 
\begin{equation}
\Lambda_{\frac{1}{2}\mathrm{arbitrary}} = \Lambda_{\frac{1}{2}6} \Lambda_{\frac{1}{2}5} \Lambda_{\frac{1}{2}4} \Lambda_{\frac{1}{2}3} \Lambda_{\frac{1}{2}2} \Lambda_{\frac{1}{2}1}
\end{equation}
Constructed this way, $\Lambda_{\frac{1}{2}\mathrm{arbitrary}}$ follows,
\begin{equation}
\Lambda_{\frac{1}{2}\mathrm{arbitrary}}^{\phantom{\frac{1}{2}\mathrm{arbitrary}}-1} \gamma^a \Lambda_{\frac{1}{2}\mathrm{arbitrary}} = \Lambda^{\phantom{\mathrm{arbitrary}}a}_{\mathrm{arbitrary}\phantom{a}b} \gamma^b
\end{equation}

If the Lagrangean density is independent of the local orientation of the coframe, then $\Psi_2$ is a solution of the curved spacetimes Dirac equation with coframe $\theta_2$: If the Lagrangean density is independent of the local orientation of the coframe, it means that for any $\theta, \Psi$,
\begin{equation}
\mathfrak{L}[\theta^{a}, \Psi] = \mathfrak{L}[\Lambda_b^{\phantom{b}a}\theta^b, \Lambda_{\frac{1}{2}}^{\phantom{\frac{1}{2}}-1}\Psi]
\end{equation}
Substituting $\theta = \theta_2, \Psi = \Psi_2$:
\begin{IEEEeqnarray}{rCl}
\mathfrak{L}[\theta_2^{\phantom{2}a}, \Psi_2] & = &  \mathfrak{L}[\Lambda_b^{\phantom{b}a}\theta_2^{\phantom{2}b}, \Lambda_{\frac{1}{2}}^{\phantom{\frac{1}{2}}-1}\Psi_2] \\
& = & \mathfrak{L}[\theta_1^{\phantom{2}a}, \Psi_1]
\end{IEEEeqnarray}
Substituting $\theta = \theta_2, \Psi = \Psi_2 + \delta \Psi_2$:
\begin{IEEEeqnarray}{rCl}
\mathfrak{L}[\theta_2, \Psi_2 + \delta \Psi_2] & = & \mathfrak{L}[\theta_1, \Lambda_{\frac{1}{2}}^{\phantom{\frac{1}{2}}-1}(\Psi_2 + \delta \Psi_2)] \\
& = & \mathfrak{L}[\theta_1, \Psi_1 + \Lambda_{\frac{1}{2}}^{\phantom{\frac{1}{2}}-1} \delta \Psi_2] \\
\mathfrak{L}[\theta_2, \Psi_2 + \delta \Psi_2] - \mathfrak{L}[\theta_2, \Psi_2] & = & \mathfrak{L}[\theta_1, \Psi_1 + \Lambda_{\frac{1}{2}}^{\phantom{\frac{1}{2}}-1} \delta \Psi_2] - \mathfrak{L}[\theta_1, \Psi_1] \IEEEeqnarraynumspace
\end{IEEEeqnarray}
Because $\Psi_1$ follows the wave equation with coframe $\theta_1$,
\begin{equation}
 \mathfrak{L}[\theta_1, \Psi_1 + \delta \Psi_1] - \mathfrak{L}[\theta_1, \Psi_1] = \mathrm{O}\big((\delta \Psi_1)^2\big) + \mathrm{d}(\ldots) 
\end{equation}
Substituting $\delta \Psi_1 = \Lambda_{\frac{1}{2}}^{\phantom{\frac{1}{2}}-1} \delta \Psi_2$:
\begin{IEEEeqnarray}{rCl}
\mathfrak{L}[\theta_1, \Psi_1 + \Lambda_{\frac{1}{2}}^{\phantom{\frac{1}{2}}-1} \delta \Psi_2] - \mathfrak{L}[\theta_1, \Psi_1] & = & \mathrm{O}\big((\Lambda_{\frac{1}{2}}^{\phantom{\frac{1}{2}}-1} \delta \Psi_2)^2\big) + \mathrm{d}(\ldots) \\
& = & \mathrm{O}\big((\delta \Psi_2)^2\big) + \mathrm{d}(\ldots) \\
\mathfrak{L}[\theta_2, \Psi_2 + \delta \Psi_2] - \mathfrak{L}[\theta_2, \Psi_2] & = & \mathrm{O}\big((\delta \Psi_2)^2\big) + \mathrm{d}(\ldots) 
\end{IEEEeqnarray}
therefore $\Psi_2$ follows the wave equation with coframe $\theta_2$.

Following is a calculation which shows that $\mathfrak{L}[\theta_2, \Psi_2] = \mathfrak{L}[\theta_1, \Psi_1]$:
\begin{IEEEeqnarray}{rCl}
\IEEEeqnarraymulticol{3}{l}{ \mathfrak{L}[\theta_2, \Psi_2] = -\frac{1}{2} \mathrm{d} A \wedge * \mathrm{d} A - e\overline{\Psi_2} \gamma^a A \wedge * \theta_{2a} \Psi_2\:{+} } \notag \\
& & {+}\:i \overline{\Psi_2} \left(\gamma^a \mathrm{d}\Psi_2 \wedge * \theta_{2a} - \frac{1}{4}\gamma^a \gamma^b \gamma^c  \omega_{2bc} \wedge *\theta_{2a} \Psi_2 + m \Psi_2 * \! 1 \right) \IEEEeqnarraynumspace \\
\qquad & = & -\frac{1}{2} \mathrm{d} A \wedge * \mathrm{d} A - e\overline{\Psi_1} \Lambda_{\frac{1}{2}}^{\phantom{\frac{1}{2}}-1} \gamma^a A(v_2^{\phantom{2}a}) \Psi_2 * \! 1\:{+} \notag \\
& & {+}\:i \overline{\Psi_1} \Lambda_{\frac{1}{2}}^{\phantom{\frac{1}{2}}-1} \left(\gamma^a \mathrm{d}\Psi_2(v_{2a}) - \frac{1}{4}\gamma^a \gamma^b \gamma^c  \omega_{2bc}(v_{2a}) \Psi_2 + m \Psi_2 \right) * \! 1 \label{eq102} \IEEEeqnarraynumspace
\end{IEEEeqnarray}
Because $\theta_1$ and $\theta_2$ represent the same metric, the Hodge dual is the same using either coframe.
\begin{multline}
\mathfrak{L}[\theta_2, \Psi_2] = -\frac{1}{2} \mathrm{d} A \wedge * \mathrm{d} A - e\overline{\Psi_1} \Lambda_{\frac{1}{2}}^{\phantom{\frac{1}{2}}-1} \gamma^a A(v_2^{\phantom{2}a}) \Lambda_{\frac{1}{2}}\Psi_1 * \! 1\:  + \\
{+}\:i \overline{\Psi_1} \Lambda_{\frac{1}{2}}^{\phantom{\frac{1}{2}}-1}\bigg( \gamma^a \Lambda_{\frac{1}{2}} \mathrm{d}\Psi_1(v_{2a}) + \gamma^a \left(v_{2a} \Lambda_{\frac{1}{2}} \right) \Psi_1 - \frac{1}{4}\gamma^a \gamma^b \gamma^c  \omega_{2bc}(v_{2a})\:{+} \\ 
{+}\:m \Lambda_{\frac{1}{2}} \Psi_1 \bigg) * \! 1 \label{eq112}
\end{multline}

The second term on the right of \eqref{eq112} is,
\begin{IEEEeqnarray}{rCl}
- e\overline{\Psi_1} \Lambda_{\frac{1}{2}}^{\phantom{\frac{1}{2}}-1} \gamma^a A(v_2^{\phantom{2}a}) \Lambda_{\frac{1}{2}}\Psi_1 * \! 1 & = & -e \overline{\Psi_1} \Lambda_{\frac{1}{2}}^{\phantom{\frac{1}{2}}-1} \gamma^a \Lambda_{\frac{1}{2}} A(v_{2a}) \Psi_1 * \! 1\\
& = & -e\overline{\Psi_1}\Lambda^a_{\phantom{a}b} \gamma^b A(v_{2a}) \Psi_1 * \! 1 \\
& = & -e\overline{\Psi_1} \gamma^a A(v_{1a}) \Psi_1 * \! 1 
\end{IEEEeqnarray}

I apply, 
\begin{equation}
\Lambda_{\frac{1}{2}}^{\phantom{\frac{1}{2}}-1} \gamma^a \Lambda_{\frac{1}{2}} = \Lambda^a_{\phantom{a}b} \gamma^b
\end{equation}
to the third term on the right of \eqref{eq112}:
\begin{IEEEeqnarray}{rCl}
i \overline{\Psi_1} \Lambda_{\frac{1}{2}}^{\phantom{\frac{1}{2}}-1} \gamma^a \Lambda_{\frac{1}{2}} (v_{2a} \Psi_1) * \! 1 & = & 
i \overline{\Psi_1} \Lambda^a_{\phantom{a}b} \gamma^b (v_{2a} \Psi_1) * \! 1 \\
& = & i \overline{\Psi_1} \Lambda^a_{\phantom{a}b} \gamma^b \Lambda_a^{\phantom{a}c} (v_{1c} \Psi_1) * \! 1 \\
& = & i \overline{\Psi_1} \gamma^a (v_{1a} \Psi_1) * \! 1
\end{IEEEeqnarray}

The fourth term on the right of \eqref{eq112} is, 
\begin{IEEEeqnarray}{rCl}
i \overline{\Psi_1}\Lambda_{\frac{1}{2}}^{\phantom{\frac{1}{2}}-1} \gamma^a \left(v_{2a} \Lambda_{\frac{1}{2}} \right) \Psi_1 * \! 1 & = &  i \overline{\Psi_1} \Lambda_{\frac{1}{2}}^{\phantom{\frac{1}{2}}-1} \gamma^a \left( v_{2a} \mathrm{exp}\left(\frac{1}{2} k_{bc} S^{bc} \right) \right) \Psi_1 * \! 1 \IEEEeqnarraynumspace \label{eq120}
\end{IEEEeqnarray} 
Because all but at most one of the $k_{bc}$ are 0 everywhere in spacetime, $\frac{1}{2} k_{bc} S^{bc}$ commutes with its derivative. A typical term of the power series expansion of ${\Lambda_{\frac{1}{2}}=\mathrm{exp}\left(\frac{1}{2} k_{bc} S^{bc} \right)}$ is:
\begin{equation}
\frac{1}{n!} \frac{1}{2^n} (k_{rs})^n (S^{rs} - S^{sr})^n
\end{equation}
Its derivative is,
\begin{multline}
\frac{1}{n!} \frac{1}{2^n} n (k_{rs})^{n - 1} (S^{rs} - S^{sr})^n v_{2a} k_{rs} = \\
= \frac{1}{(n - 1)!} \frac{1}{2^{n-1}} (k_{rs})^{n - 1}(S^{rs} - S^{sr})^{n - 1} \frac{1}{2}(S^{rs} - S^{sr}) v_{2a} k_{rs}
\end{multline}
So the derivative of $\Lambda_{\frac{1}{2}}$ is,
\begin{equation}
v_{2a} \Lambda_{\frac{1}{2}} = \frac{1}{2} \Lambda_{\frac{1}{2}} S^{bc} v_{2a} k_{bc}
\end{equation}
The right side of \eqref{eq120} becomes,
\begin{IEEEeqnarray}{rCl}
\IEEEeqnarraymulticol{3}{l}{ \frac{i}{2} \overline{\Psi_1} \Lambda_{\frac{1}{2}}^{\phantom{\frac{1}{2}}-1} \gamma^a \Lambda_{\frac{1}{2}}S^{bc} \left( v_{2a} k_{bc} \right) \Psi_1 * \! 1 = \qquad \qquad \qquad  \qquad \qquad } \IEEEeqnarraynumspace \\
\qquad \qquad & = & \frac{i}{2} \overline{\Psi_1} \Lambda^a_{\phantom{a}d} \gamma^d S^{bc} \left( v_{2a} k_{bc} \right) \Psi_1 * \! 1 \\
& = & \frac{i}{2} \overline{\Psi_1} \gamma^a S^{bc} \left( v_{1a} k_{bc} \right) \Psi_1 * \! 1
\end{IEEEeqnarray}

The fifth term on the right of \eqref{eq112} is,
\begin{equation}
 - i \overline{\Psi_1} \Lambda_{\frac{1}{2}}^{\phantom{\frac{1}{2}}-1} \frac{1}{4} \gamma^a \gamma^b \gamma^c \omega_{2bc}(v_{2a}) \Lambda_{\frac{1}{2}} \Psi_1 * \! 1
\end{equation}
with,
\begin{equation}
\omega_{2bc}(v_{2a}) = \frac{1}{2} \Big( \mathrm{d}\theta_{2a}(v_{2b}, v_{2c}) + \mathrm{d}\theta_{2b}(v_{2a}, v_{2c}) - \mathrm{d}\theta_{2c}(v_{2a}, v_{2b}) \Big) \label{eq128}
\end{equation}
Raising the index $a$, the first term inside the parentheses on the right of \eqref{eq128} is,
\begin{IEEEeqnarray}{rCl}
\mathrm{d}\theta_2^{\phantom{2}a}(v_{2b}, v_{2c}) & = & \mathrm{d} (\Lambda^a_{\phantom{a}f} \theta_{1}^{\phantom{1}f} )(v_{2b}, v_{2c}) \\
 & = & \mathrm{d} \Big(\big(\mathrm{exp}(k^d_{\phantom{d}e}\mathfrak{I}_d^{\phantom{d}e}) \big)^a_{\phantom{a}f}\theta_{1}^{\phantom{1}f}\Big)(v_{2b}, v_{2c}) \\
 & = & \Big( v_{1g} \big(\mathrm{exp}(k^d_{\phantom{d}e}\mathfrak{I}_d^{\phantom{d}e}) \big)^a_{\phantom{a}f} \Big) (\theta_1^{\phantom{1}g} \wedge \theta_{1}^{\phantom{1}f} ) (v_{2b}, v_{2c})\:{+} \notag \\
& & {+}\:\Lambda^a_{\phantom{a}f} \mathrm{d}\theta_{1}^{\phantom{1}f}\left(v_{2b}, v_{2c}\right) \label{eq153}
\end{IEEEeqnarray}
The first term on the right of \eqref{eq153} is,
\begin{multline}
\Big( v_{1g} \big(\mathrm{exp}(k^d_{\phantom{d}e}\mathfrak{I}_d^{\phantom{d}e}) \big)^a_{\phantom{a}f} \Big) (\theta_1^{\phantom{1}g} \wedge \theta_{1}^{\phantom{1}f} ) (v_{2b}, v_{2c}) = \\
 =  \big(v_{1g} \mathrm{exp}(k^d_{\phantom{d}e}\mathfrak{I}_d^{\phantom{d}e}) \big)^a_{\phantom{a}f} (\theta_1^{\phantom{1}g} \wedge \theta_{1}^{\phantom{1}f})(v_{2b}, v_{2c}) \label{eq88}
\end{multline}
Because all but at most one of the $k_{de}$ are 0 everywhere in spacetime, $k_{de}\mathfrak{I}^{de}$ commutes with its derivative. A typical term of the power series expansion of  $\Lambda = \mathrm{exp}\left(k_{de}\mathfrak{I}^{de}\right)$ is,
\begin{equation}
\frac{1}{n!}(k_{rs})^n (\mathfrak{I}^{rs} - \mathfrak{I}^{sr})^n
\end{equation}
Its derivative is,
\begin{multline}
\frac{1}{n!} n (k_{rs})^{n - 1} (\mathfrak{I}^{rs} - \mathfrak{I}^{sr})^n v_{1g} k_{rs} = \\
= \frac{1}{(n - 1)!} (k_{rs})^{n - 1}(\mathfrak{I}^{rs} - \mathfrak{I}^{sr})^{n - 1} (\mathfrak{I}^{rs} - \mathfrak{I}^{sr}) v_{1g} k_{rs}
\end{multline}
So the derivative of $\Lambda$ is,
\begin{equation}
v_{1g} \Lambda = \Lambda \mathfrak{I}^{de} v_{1g} k_{de}
\end{equation}
So the right side of \eqref{eq88} becomes,
\begin{IEEEeqnarray}{rCl}
\IEEEeqnarraymulticol{3}{l}{ \Big( \mathrm{exp} (k^d_{\phantom{d}e} \mathfrak{I}_d^{\phantom{d}e}) \mathfrak{I}_h^{\phantom{h}i} v_{1g} k^h_{\phantom{h}i} \Big)^a_{\phantom{a}f}(\theta_1^{\phantom{1}g} \wedge \theta_{1}^{\phantom{1}f})(v_{2b}, v_{2c}) } \notag \\
\qquad & = & \Big( \mathrm{exp} (k^d_{\phantom{d}e} \mathfrak{I}_d^{\phantom{d}e} ) \mathfrak{I}_h^{\phantom{h}i} \Big)^a_{\phantom{a}f} (v_{1g} k^h_{\phantom{h}i} )(\theta_1^{\phantom{1}g} \wedge \theta_{1}^{\phantom{1}f})(v_{2b}, v_{2c}) \\
& = & \Lambda^a_{\phantom{a}h} \delta^i_{\phantom{i}f} (v_{1g} k^h_{\phantom{h}i} )( \theta_1^{\phantom{1}g} \wedge \theta_{1}^{\phantom{1}f})(v_{2b}, v_{2c}) \\
& = & \Lambda^a_{\phantom{a}h} (v_{1g} k^h_{\phantom{h}f} ) ( \theta_1^{\phantom{1}g} \wedge \theta_{1}^{\phantom{1}f}) (v_{2b}, v_{2c}) \\
& = & \Lambda^a_{\phantom{a}h} (v_{1g} k^h_{\phantom{h}f} ) \Big(\theta_1^{\phantom{1}g}(v_{2b})\theta_{1}^{\phantom{1}f}(v_{2c}) - \theta_{1}^{\phantom{1}g}(v_{2c})\theta_{1}^{\phantom{1}f}(v_{2b})\Big) \\
& = & \Lambda^a_{\phantom{a}h} (v_{1g} k^h_{\phantom{h}f} )\Big(\theta_1^{\phantom{1}g}(\Lambda_b^{\phantom{b}j}v_{1j})\theta_{1}^{\phantom{1}f} ( \Lambda_c^{\phantom{c}k}v_{1k} ) - \theta_{1}^{\phantom{1}g}(\Lambda_c^{\phantom{c}k}v_{1k})\theta_{1}^{\phantom{1}f}(\Lambda_b^{\phantom{b}j}v_{1j})\Big) \IEEEeqnarraynumspace \\
& = & \Lambda^a_{\phantom{a}h} (v_{1g} k^h_{\phantom{h}f} )\Big(\Lambda_b^{\phantom{b}j} \delta^g_{\phantom{g}j} \Lambda_c^{\phantom{c}k}\delta^f_{\phantom{f}k} - \Lambda_c^{\phantom{c}k}\delta^g_{\phantom{g}k} \Lambda_b^{\phantom{b}j}\delta^f_{\phantom{f}j}\Big) \\
& = & \Lambda^a_{\phantom{a}h}\Lambda_b^{\phantom{b}g} \Lambda_c^{\phantom{c}f}\Big(v_{1g} k^h_{\phantom{h}f} - v_{1f} k^h_{\phantom{h}g} \Big) \\
& = & \Lambda^a_{\phantom{a}f}\Lambda_b^{\phantom{b}g} \Lambda_c^{\phantom{c}h}\Big(v_{1g} k^f_{\phantom{f}h} - v_{1h} k^f_{\phantom{f}g} \Big)
\end{IEEEeqnarray}
The second term on the right of \eqref{eq153} is,
\begin{equation}
\Lambda^a_{\phantom{a}f} \mathrm{d}\theta_{1}^{\phantom{1}f}\left(v_{2b}, v_{2c}\right) = \Lambda^a_{\phantom{a}f} \Lambda_b^{\phantom{b}g} \Lambda_c^{\phantom{c}h} \mathrm{d}\theta_{1}^{\phantom{1}f}\left(v_{1g}, v_{1h}\right)
\end{equation}
Lowering the index $a$, I get,
\begin{multline}
\mathrm{d}\theta_{2a}(v_{2b}, v_{2c}) = \Lambda_a^{\phantom{a}f} \Lambda_b^{\phantom{b}g} \Lambda_c^{\phantom{c}h} \mathrm{d}\theta_{1f}\left(v_{1g}, v_{1h}\right)\:{+} \\ 
{+}\:\Lambda_a^{\phantom{a}f}\Lambda_b^{\phantom{b}g} \Lambda_c^{\phantom{c}h}\big(v_{1g} k_{fh} - v_{1h} k_{fg} \big)
\end{multline}
Similarly, 
\begin{multline}
\omega_{2bc}(v_{2a}) = \Lambda_a^{\phantom{a}f} \Lambda_b^{\phantom{b}g} \Lambda_c^{\phantom{c}h} \bigg( \omega_{1gh}(v_{1f}) + \frac{1}{2} \Big( v_{1g} k_{fh} - v_{1h} k_{fg}\:{+} \\
{+}\:v_{1f} k_{gh} - v_{1h} k_{gf} - v_{1f} k_{hg} + v_{1g} k_{hf} \Big)\bigg) 
\end{multline}
and because $k$ is antisymmetric,
\begin{equation}
\omega_{2bc}(v_{2a}) = \Lambda_a^{\phantom{a}f} \Lambda_b^{\phantom{b}g} \Lambda_c^{\phantom{c}h} \Big(\omega_{1gh}(v_{1f}) + v_{1f} k_{gh} \Big) 
\end{equation}
The fifth term on the right of \eqref{eq112} becomes,
\begin{multline}
 - \frac{1}{4} \overline{\Psi_1} \Lambda_{\frac{1}{2}}^{\phantom{\frac{1}{2}}-1}  \gamma^a \gamma^b \gamma^c \omega_{2bc}(v_{2a}) \Lambda_{\frac{1}{2}} \Psi_1 =  \\
 = - \frac{1}{4} \overline{\Psi_1} \Lambda_{\frac{1}{2}}^{\phantom{\frac{1}{2}}-1}  \gamma^a \gamma^b \gamma^c \Lambda_{\frac{1}{2}} \Lambda_a^{\phantom{a}f} \Lambda_b^{\phantom{b}g} \Lambda_c^{\phantom{c}h} \Big(\omega_{1gh}(v_{1f}) + v_{1f} k_{gh} \Big) \Psi_1
\end{multline}
Using that,
\begin{equation}
\Lambda_a^{\phantom{a}b} \gamma^a = \Lambda_{\frac{1}{2}} \gamma^b \Lambda_{\frac{1}{2}}^{\phantom{\frac{1}{2}}-1}
\end{equation}
The fifth term on the right of \eqref{eq112} becomes,
\begin{IEEEeqnarray}{rCl}
\IEEEeqnarraymulticol{3}{l}{
 - \frac{i}{4} \overline{\Psi_1}  \Lambda_{\frac{1}{2}}^{\phantom{\frac{1}{2}}-1}  \gamma^a \gamma^b \gamma^c \omega_{2bc}(v_{2a}) \Lambda_{\frac{1}{2}} \Psi_1 * \! 1 = } \notag \\
 \qquad \qquad & = & {-}\:\frac{i}{4} \overline{\Psi_1}  \Lambda_{\frac{1}{2}}^{\phantom{\frac{1}{2}}-1} \Lambda_{\frac{1}{2}} \gamma^a \Lambda_{\frac{1}{2}}^{\phantom{\frac{1}{2}}-1}\Lambda_{\frac{1}{2}}\gamma^b\Lambda_{\frac{1}{2}}^{\phantom{\frac{1}{2}}-1}\Lambda_{\frac{1}{2}} \gamma^c  \Lambda_{\frac{1}{2}}^{\phantom{\frac{1}{2}}-1}\Lambda_{\frac{1}{2}}\:{\times} \IEEEeqnarraynumspace \notag \\
 &  & {\times}\:\Big(\omega_{1bc}(v_{1a}) + v_{1a} k_{bc} \Big) \Psi_1 * \! 1 \\
 & = & {-}\:\frac{i}{4} \overline{\Psi_1} \gamma^a \gamma^b \gamma^c  \Big(\omega_{1bc}(v_{1a}) + v_{1a} k_{bc} \Big) \Psi_1 * \! 1 
 \end{IEEEeqnarray}

Gathering terms, \eqref{eq112} becomes,
\begin{IEEEeqnarray}{rCl}
\mathfrak{L}[\theta_2, \Psi_2] & = & -\frac{1}{2} \mathrm{d} A \wedge * \mathrm{d} A - e\overline{\Psi_1} \gamma^a A(v_{1a}) \Psi_1 * \! 1\:{+} \notag \\
& & {+}\:i \overline{\Psi_1} \bigg( \gamma^a v_{1a} \Psi_1 + \frac{1}{2}\gamma^a S^{bc} (v_{1a} k_{bc}) \Psi_1 \:{-} \notag \\ 
& & {-}\:\frac{1}{4} \gamma^a \gamma^b \gamma^c  \Big(\omega_{1bc}(v_{1a}) + v_{1a} k_{bc} \Big) \Psi_1 + m \Psi_1 \bigg) * \! 1 \\
& = &  -\frac{1}{2} \mathrm{d} A \wedge * \mathrm{d} A - e\overline{\Psi_1} \gamma^a A(v_{1a}) \Psi_1 * \! 1\:{+} \notag \\
& & {+}\:i \overline{\Psi_1} \left( \gamma^a v_{1a} \Psi_1 - \frac{1}{4} \gamma^a \gamma^b \gamma^c  \omega_{1bc}(v_{1a}) \Psi_1 + m  \Psi_1\right) * \! 1 \IEEEeqnarraynumspace \label{eq142} \\
& = & \mathfrak{L}[\theta_1, \Psi_1] 
 \end{IEEEeqnarray}
This shows that the Lagrangean density is independent of the local orientation of the coframe.

A second proof that the Dirac equation in curved spacetimes is invariant under a local rotation of the coframe is by re-reading \eqref{eq102} - \eqref{eq142} while ignoring the first electromagnetic term and for the other terms, ignoring the $i\overline{\Psi_1}$ in front and the $*1$ after. Then \eqref{eq102} - \eqref{eq142} can be summarized as,
\begin{multline}
\Lambda_{\frac{1}{2}}^{\phantom{\frac{1}{2}}-1} \left(\gamma^a \mathrm{d}\Psi_2(v_{2a}) +i e \gamma^a A(v_2^{\phantom{2}a}) \Psi_2 - \frac{1}{4}\gamma^a \gamma^b \gamma^c  \omega_{2bc}(v_{2a}) \Psi_2 + m \Psi_2 \right) = \\
= \left( \gamma^a v_{1a} \Psi_1 + i e \gamma^a A(v_{1a})\Psi_1 - \frac{1}{4} \gamma^a \gamma^b \gamma^c  \omega_{1bc}(v_{1a})\Psi_1 + m \Psi_1\right) 
\end{multline}

\section{Conservation of current} \label{Conservation of current}

I now calculate,
\begin{IEEEeqnarray}{rCl}
\mathrm{d} \Psi \wedge *\theta_0 & = & \mathrm{d} \Psi \wedge \theta^1 \wedge \theta^2 \wedge \theta^3 \\
& = & \mathrm{d} \Psi(v_0) \theta^0 \wedge \theta^1 \wedge \theta^2 \wedge \theta^3 \\
& = & \mathrm{d} \Psi(v_0) * \! 1
\end{IEEEeqnarray}
Therefore,
\begin{IEEEeqnarray}{rCl}
* \left( \mathrm{d} \Psi \wedge *\theta_0 \right) & = & * \left( \mathrm{d} \Psi(v_0) * \! 1 \right) \\
& = & \mathrm{d} \Psi(v_0) * \! * 1\\
& = & -\mathrm{d} \Psi(v_0)
\end{IEEEeqnarray}
Generally,
\begin{equation}
* \left( \mathrm{d} \Psi \wedge *\theta_a \right) = -\mathrm{d} \Psi(v_a)
\end{equation}
So,
\begin{equation}
v_a \Psi = - * \left( \mathrm{d} \Psi \wedge *\theta_a \right)
\end{equation}
Substituting $v_a \Psi = - * \left( \mathrm{d} \Psi \wedge *\theta_a \right)$ in the Dirac equation in curved spacetimes, I get,
\begin{equation}
 - \gamma^a *\left( \mathrm{d} \Psi \wedge *\theta_a \right) + ie\gamma^a A(v_a) \Psi- \frac{1}{4} \gamma^a \gamma^b \gamma^c \omega_{bc}(v_a) \Psi + m \Psi = 0 \label{eq164}
\end{equation}
Its Hermitean conjugate is,
\begin{equation}
 - *\left( \mathrm{d} \Psi^\dagger \wedge *\theta_a \right) \gamma^{a\dagger} - ie \Psi^\dagger \gamma^{a\dagger} A(v_a) - \frac{1}{4}  \omega_{bc}(v_a) \Psi^\dagger \gamma^{c\dagger} \gamma^{b\dagger} \gamma^{a\dagger} + m \Psi^\dagger = 0
\end{equation}
Multiplying by $\gamma^0$ on the right,
\begin{IEEEeqnarray}{rCl}
\IEEEeqnarraymulticol{3}{l}{- *\left( \mathrm{d} \Psi^\dagger \wedge *\theta_a \right) \gamma^{a\dagger} \gamma^0 - ie \Psi^\dagger \gamma^{a\dagger} \gamma^0 A(v_a) - \frac{1}{4}  \omega_{bc}(v_a) \Psi^\dagger \gamma^{c\dagger} \gamma^{b\dagger} \gamma^{a\dagger} \gamma^0\:{+} }\notag \\
{+}\:m \Psi^\dagger \gamma^0 & = & 0 \\
\IEEEeqnarraymulticol{3}{l}{*\left( \mathrm{d} \Psi^\dagger \wedge *\theta_a \right) \gamma^0 \gamma^{a}  + ie \Psi^\dagger \gamma^0\gamma^{a} A(v_a)+ \frac{1}{4}  \omega_{bc}(v_a) \Psi^\dagger \gamma^0 \gamma^{c} \gamma^{b} \gamma^{a}\:{+} }\notag \\
{+}\: m \Psi^\dagger \gamma^0 & = & 0 \\
*\left( \mathrm{d} \overline{\Psi} \wedge *\theta_a \right)  \gamma^{a}  + ie \overline{\Psi} \gamma^{a} A(v_a) + \frac{1}{4}  \omega_{bc}(v_a) \overline{\Psi} \gamma^{c} \gamma^{b} \gamma^{a}  + m \overline{\Psi} & = & 0 \label{eq168}
\end{IEEEeqnarray}

The current is,
\begin{equation}
j = - e\overline{\Psi} \gamma^a \Psi * \! \theta_{a}
\end{equation}
Conservation of current means,
\begin{IEEEeqnarray}{rCl}
* \mathrm{d}j & = & 0 \\
 - e* \left( \mathrm{d}\overline{\Psi} \wedge *\theta_a \right) \gamma^a \Psi -
 e\overline{\Psi} \gamma^a * \left( \mathrm{d}\Psi \wedge *\theta_a \right) -
 e\overline{\Psi} \gamma^a \Psi * \! \mathrm{d} \! * \!\theta_{a} & = & 0 \IEEEeqnarraynumspace
 \end{IEEEeqnarray}
Using \eqref{eq164} and \eqref{eq168},
\begin{equation}
 \frac{e}{4}  \omega_{bc}(v_a) \overline{\Psi} \left(\gamma^a \gamma^b \gamma^c + \gamma^{c} \gamma^{b} \gamma^{a} \right) \Psi - e\overline{\Psi} \gamma^a \Psi * \! \mathrm{d} \! * \!\theta_{a} 
 = 0 \label{eq104}
 \end{equation}
The products of three different gammas cancel leaving,
\begin{IEEEeqnarray}{rCl}
\frac{e}{4} \overline{\Psi} \left(4 \gamma^c \omega^a_{\phantom{a}c}(v_a) \right) \Psi - e\overline{\Psi} \gamma^a \Psi * \! \mathrm{d} \! * \!\theta_{a} & = & 0  \\
 e\overline{\Psi} \left(\gamma^b \omega^a_{\phantom{a}b}(v_a) \right) \Psi - e\overline{\Psi} \gamma^a \Psi * \! \mathrm{d} \! * \!\theta_{a} & = & 0 \\  
\IEEEeqnarraymulticol{3}{c}{\omega^a_{\phantom{a}b}(v_a) = -\mathrm{d}\theta^a(v_b, v_a) } \label{eq153b} \\ 
- e\overline{\Psi} \gamma^b \Psi \, \mathrm{d}\theta^a(v_b, v_a) - e\overline{\Psi} \gamma^a \Psi * \! \mathrm{d} \! * \!\theta_{a} & = & 0  
\end{IEEEeqnarray}
For the last term,
\begin{IEEEeqnarray}{rCl}
\IEEEeqnarraymulticol{3}{c}{*\theta_{0} = \theta^1 \wedge \theta^2 \wedge \theta^3 } \\
\IEEEeqnarraymulticol{3}{c}{\mathrm{d} \! * \!\theta_{0} = \mathrm{d}\theta^1 \wedge \theta^2 \wedge \theta^3 
- \theta^1 \wedge \mathrm{d} \theta^2 \wedge \theta^3 
+ \theta^1 \wedge \theta^2 \wedge \mathrm{d} \theta^3 } \\
*\left(\mathrm{d}\theta^1 \wedge \theta^2 \wedge \theta^3 \right) & = &
*\left( \mathrm{d}\theta^1 (v_0, v_1) \theta^0 \wedge \theta^1 \wedge \theta^2 \wedge \theta^3 \right) \\
& = & *\left( \mathrm{d}\theta^1 (v_0, v_1) * \! 1 \right) \\ 
& = & \mathrm{d}\theta^1 (v_0, v_1) * \! * 1 \\
& = & - \mathrm{d}\theta^1 (v_0, v_1)
\end{IEEEeqnarray}
Similarly,
\begin{equation}
* \mathrm{d} \! * \! \theta_{a} = - \mathrm{d}\theta^b (v_a, v_b) \label{eq115}
\end{equation}
And the equation for the conservation of current becomes,
\begin{IEEEeqnarray}{rCl}
* \mathrm{d} j & = & - e\overline{\Psi} \gamma^a \Psi \, \mathrm{d}\theta^b(v_a, v_b) + e\overline{\Psi} \gamma^a \Psi \, \mathrm{d}\theta^b (v_a, v_b) \\
& = & 0
\end{IEEEeqnarray}

\section{Alternative formulation} \label{Alternative formulation}

The term,
\begin{equation}
\omega_{bc}(v_a)
\end{equation}
can be written in an alternative formulation:
\begin{align}
\omega_{bc}(v_a)  = & \:\frac{1}{2} \Big( \mathrm{d}\theta_a(v_b, v_c) + \mathrm{d}\theta_b(v_a, v_c) - \mathrm{d}\theta_c(v_a, v_b) \Big) \\
= & \: \frac{1}{2} \Big( \mathrm{d}\theta_a(v_b^{\phantom{b}\nu} \partial_\nu, v_c^{\phantom{c}\rho} \partial_\rho) + \mathrm{d}\theta_b(v_a^{\phantom{a}\mu} \partial_\mu, v_c^{\phantom{c}\rho} \partial_\rho) - \mathrm{d}\theta_c(v_a^{\phantom{a}\mu} \partial_\mu, v_b^{\phantom{b}\nu} \partial_\nu) \Big) \\
= & \:\frac{1}{2} \Big( v_b^{\phantom{b}\nu} v_c^{\phantom{c}\rho} \big(\partial_\nu \theta_{a \rho} - \partial_\rho \theta_{a \nu} \big) 
+ v_a^{\phantom{a}\mu} v_c^{\phantom{c}\rho} \big( \partial_\mu \theta_{b \rho} - \partial_\rho \theta_{b \mu} \big)\:{+} \notag \\
& \:{+}\:v_a^{\phantom{a}\mu} v_b^{\phantom{b}\nu} \big( - \partial_\mu\theta_{c \nu} + \partial_\nu \theta_{c \mu} \big) \Big) \label{eq189}
\end{align}
Writing the Christoffel gamma as,
\begin{equation}
\Gamma_{\rho \mu \nu } = \frac{1}{2}\Big( \partial_\mu g_{\rho \nu} + \partial_\nu g_{\rho \mu} - \partial_\rho g_{\nu \mu} \Big)
\end{equation}
and using that,
\begin{equation}
g_{\mu \nu} = \eta_{a b} \theta^{a}_{\phantom{a}\mu} \theta^{b}_{\phantom{b}\nu} \end{equation}
the Christoffel gamma becomes,
\begin{IEEEeqnarray}{rCl}
\Gamma_{\rho \mu \nu} & = & \frac{1}{2}\Big( \eta_{d e} \partial_\mu  (\theta^{d}_{\phantom{d}\rho} \theta^{e}_{\phantom{e}\nu}) + \eta_{d e} \partial_\nu (\theta^{d}_{\phantom{d}\rho} \theta^{e}_{\phantom{e}\mu}) - \eta_{d e}\partial_\rho ( \theta^{d}_{\phantom{d}\nu} \theta^{e}_{\phantom{e}\mu}) \Big) \\
\IEEEeqnarraymulticol{3}{l}{ =  \frac{1}{2} \Big( \theta^{d}_{\phantom{d}\nu} \partial_\mu  \theta_{d \rho} + \theta^{d}_{\phantom{d}\rho} \partial_\mu \theta_{d \nu} + \theta^{d}_{\phantom{d}\mu} \partial_\nu \theta_{d \rho} + \theta^{d}_{\phantom{d}\rho} \partial_\nu \theta_{d \mu} - \theta^{d}_{\phantom{e}\mu} \partial_\rho \theta_{d \nu}  - \theta^{d}_{\phantom{d}\nu} \partial_\rho \theta_{d\mu} \Big) } \IEEEeqnarraynumspace \\
 v_a^{\phantom{a} \mu} v_b^{\phantom{b} \nu} v_c^{\phantom{c} \rho}\Gamma_{\rho \mu \nu} & = & \frac{1}{2} \Big( v_a^{\phantom{a} \mu} v_c^{\phantom{c} \rho} \big(\partial_\mu  \theta_{b \rho} - \partial_\rho \theta_{b\mu} \big) + v_a^{\phantom{a} \mu} v_b^{\phantom{b} \nu} \big( \partial_\mu \theta_{c \nu} + \partial_\nu \theta_{c \mu} \big)\:{+} \notag \\
& & {+}\:v_b^{\phantom{b} \nu} v_c^{\phantom{c} \rho} \big( \partial_\nu \theta_{a \rho} - \partial_\rho \theta_{a \nu} \big) \Big)
 \end{IEEEeqnarray}
Inserting into \eqref{eq189}, I get,
\begin{equation}
\omega_{bc}(v_a) = v_a^{\phantom{a}\mu} v_b^{\phantom{b}\nu} \left( - \partial_\mu\theta_{c \nu} + \theta_{c  \rho} \Gamma^\rho_{\mu \nu} \right)
\end{equation}
An alternative formulation for the Dirac equation in curved spacetimes is,
\begin{equation}
\left(\gamma^a v_a^{\phantom{a}\mu} \partial_\mu + ie\gamma^a A (v_a) - \frac{1}{2} \gamma^a S^{bc} v_a^{\phantom{a} \mu} v_b^{\phantom{b} \nu}\left( - \partial_\mu\theta_{c \nu} + \theta_{c  \rho} \Gamma^\rho_{\mu \nu}\right) + m \right) \Psi = 0 
\end{equation}

\section{Stress-energy tensor} \label{Stress-energy tensor}

In this section, the stress-energy tensor of the electromagnetic and Dirac fields is calculated.

The Lagrangean density is,
\begin{IEEEeqnarray}{rCl}
\mathfrak{L} & = & \mathfrak{L}_{\mathrm{em}} + \mathfrak{L}_{\mathrm{int}} + \mathfrak{L}_{\mathrm{Dirac}} \\
& = & -\frac{1}{2} \mathrm{d} A \wedge * \mathrm{d} A - e\overline{\Psi} \gamma^a A \wedge * \theta_a \Psi\:{+} \notag \\
& & {+}\:i \overline{\Psi} \left(\gamma^a \mathrm{d}\Psi \wedge * \theta_a - \frac{1}{4}\gamma^a \gamma^b \gamma^c  \omega_{bc} \wedge * \theta_a \Psi + m \Psi * \! 1 \right)
\end{IEEEeqnarray} 
To calculate the stress-energy tensor, I use the same method as in coframe gravity \cite{Itin}. This consists of setting up a small variation of the coframe which vanishes on the region boundary. The variation in the Lagrangean density is,
\begin{multline}
\delta \mathfrak{L} =  -\frac{1}{2} \delta (\mathrm{d} A \wedge * \mathrm{d} A) - e\overline{\Psi}\gamma^a \Psi A \wedge \delta \! * \!\theta_a + i\overline{\Psi} \bigg(\gamma^a \mathrm{d} \Psi \wedge \delta \! * \! \theta_a \:{-}  \\
 {-}\:\frac{1}{4}\gamma^a \gamma^b \gamma^c  \Big(\delta \big( \omega_{bc}(v_a) * \! 1 \big) \Big)  \Psi + m \Psi \delta \! * \!  1\bigg) \label{eq187}
\end{multline}

The first term on the right of \eqref{eq187} is,
\begin{equation}
-\frac{1}{2} \delta (\mathrm{d} A \wedge * \mathrm{d} A)
\end{equation}
Denoting,
\begin{equation}
F_{ab} = \mathrm{d}A(v_a, v_b)
\end{equation}
the first term can be written as,
\begin{IEEEeqnarray}{rCl}
\IEEEeqnarraymulticol{3}{l}{- \frac{1}{2} \delta (\mathrm{d} A \wedge * \mathrm{d} A) = -\frac{1}{4} \delta (F_{ab} \eta^{ac}\eta^{bd}F_{cd} * \! 1)} \\
\qquad \qquad & = & - \frac{1}{2}F^{ab} \delta (F_{ab} * \! 1) + \frac{1}{4} F^{ab}F_{ab}\delta \! * \! 1 \\
\delta \! * \! 1 & = & \delta ( \theta^0 \wedge \theta^1 \wedge \theta^2 \wedge \theta^3 ) \\
& = & \delta \theta^0 \wedge \theta^1 \wedge \theta^2 \wedge \theta^3 + \theta^0 \wedge \delta \theta^1 \wedge \theta^2 \wedge \theta^3 + \ldots \\
& = & \delta \theta^m \wedge * \theta_m \\
\IEEEeqnarraymulticol{3}{l}{- \frac{1}{2} \delta (\mathrm{d} A \wedge * \mathrm{d} A) = -\frac{1}{2}F^{ab}\delta \left(\mathrm{d} A \wedge \frac{\epsilon_{abmd}}{2} \theta^m \wedge \theta^d \right)\:{+} } \\
& & \qquad \: \: \: \: \, {+}\:\frac{1}{4} F^{ab}F_{ab} \delta \theta_m \wedge * \theta^m \\
& = & \delta \theta_m \wedge \left(-\frac{1}{4}F^{ab} \mathrm{d} A \wedge \epsilon_{abcd} \eta^{cm} \theta^d + \frac{1}{4} F^{ab}F_{ab} * \theta^m \right) \\
& = & \delta \theta_m \wedge \left(-\frac{1}{4}F^{ab} F_{ef}\epsilon_{abcd} \eta^{cm} \epsilon^{nefd} + \frac{1}{4} \eta^{mn} F^{ab}F_{ab} \right) * \! \theta_n \IEEEeqnarraynumspace \\
& = & \delta \theta_m \wedge \left(-\frac{1}{4}F^{ab} F_{ef}\epsilon_{abcd} \eta^{cm} \epsilon^{efnd} + \frac{1}{4} \eta^{mn} F^{ab}F_{ab} \right) * \! \theta_n \\
& = & \delta \theta_m \wedge \left(F^{ma} F^n_{\phantom{n}a} - \frac{1}{4} \eta^{mn} F^{ab}F_{ab} \right) * \! \theta_n 
\end{IEEEeqnarray}

The second term on the right of \eqref{eq187} is,
\begin{IEEEeqnarray}{rCl}
-e\overline{\Psi} \gamma^a A \wedge \delta \! * \!  \theta_a & = & -e \overline{\Psi} \gamma^a \Psi A \wedge \delta \left( \frac{\epsilon_{abcd}}{6} \theta^b \wedge \theta^c \wedge \theta^d \right) \\*
\delta ( \epsilon_{abcd} \theta^b \wedge \theta^c \wedge \theta^d ) & = & \epsilon_{abcd} \big(\delta \theta^b \wedge \theta^c \wedge \theta^d +  \theta^b \wedge \delta \theta^c \wedge \theta^d\:{+} \notag \\
& & {+}\:\theta^b \wedge \theta^c \wedge \delta \theta^d\big) \\
\epsilon_{abcd}\theta^b \wedge \delta \theta^c \wedge \theta^d & = & - \epsilon_{abcd} \delta \theta^c \wedge  \theta^b \wedge \theta^d \\
& = & \epsilon_{acbd} \delta \theta^c \wedge  \theta^b \wedge \theta^d \\
& = & \epsilon_{abcd} \delta \theta^b \wedge  \theta^c \wedge \theta^d \\
-e\overline{\Psi} \gamma^a \Psi A \wedge \delta \! * \!  \theta_a & = & -\frac{1}{2}e \overline{\Psi} \gamma^a \Psi A \wedge \epsilon_{abcd} \delta\theta^b \wedge \theta^c \wedge \theta^d \\
& = & \delta\theta^b \wedge  \frac{1}{2} e \overline{\Psi} \gamma^a \Psi A \wedge \epsilon_{abcd} \theta^c \wedge \theta^d \\
& = & \delta\theta^m \wedge  \frac{1}{2} e \overline{\Psi} \gamma^a \Psi A(v_e) \epsilon_{amcd} \epsilon^{necd} *\theta_n \\
& = & \delta\theta_m \wedge  -\frac{1}{2} e \overline{\Psi} \gamma^a \Psi A(v_e)  \epsilon_{agcd} \eta^{gm} \epsilon^{encd} *\theta_n \\
& = & \delta\theta_m \wedge -  e \Big( \eta^{mn}\overline{\Psi}\gamma^a \Psi A(v_a) - \overline{\Psi} \gamma^n \Psi A(v^m) \Big) * \! \theta_n \IEEEeqnarraynumspace
\end{IEEEeqnarray}

Similarly, the third term on the right of \eqref{eq187} is,
\begin{equation}
i\overline{\Psi} \gamma^a \mathrm{d}\Psi \wedge \delta \! * \! \theta_a = 
\delta\theta_m \wedge i \Big( \eta^{mn}\overline{\Psi} \gamma^a \mathrm{d}  \Psi(v_a) - \overline{\Psi} \gamma^n \mathrm{d}  \Psi(v^m) \Big) * \! \theta_n \label{eq206}
\end{equation}
Using that,
\begin{IEEEeqnarray}{rCl}
\gamma^a \gamma^b \gamma^c \mathrm{d}\theta_b(v_a, v_c) & = & \gamma^b \gamma^a \gamma^c \mathrm{d}\theta_a(v_b, v_c) \\
- \gamma^a \gamma^b \gamma^c \mathrm{d}\theta_c(v_a, v_b) & = & - \gamma^c \gamma^b \gamma^a \mathrm{d}\theta_a(v_c, v_b) \\
& = &  - \gamma^b \gamma^c \gamma^a \mathrm{d}\theta_a(v_b, v_c) \\
& = & \big( - 2 \eta^{ac} \gamma^b + \gamma^b \gamma^a \gamma^c \big) \mathrm{d}\theta_a(v_b, v_c)
\end{IEEEeqnarray}
the fourth term on the right of \eqref{eq187} is,
\begin{IEEEeqnarray}{rCl}
\IEEEeqnarraymulticol{3}{l}{- \frac{i}{4}\overline{\Psi}\gamma^a \gamma^b \gamma^c  \Big( \delta \big( \omega_{bc}(v_a) * \! 1 \big) \Big)  \Psi = } \notag\\
\qquad \qquad & = & - \frac{i}{8}\overline{\Psi} \big( \gamma^a \gamma^b \gamma^c + \gamma^b \gamma^a \gamma^c - 2 \eta^{ac} \gamma^b + \gamma^b \gamma^a \gamma^c \big)\:{\times} \notag \\
& & {\times}\:\Big( \delta \big( \mathrm{d}\theta_a(v_b, v_c) * \! 1 \big) \Big)  \Psi \\
& = & - \frac{i}{8}\overline{\Psi} \big( 2\eta^{ab} \gamma^c - 2 \eta^{ac} \gamma^b + \gamma^b \gamma^a \gamma^c \big) \Big( \delta \big( \mathrm{d}\theta_a(v_b, v_c)  * \! 1 \big) \Big)  \Psi \IEEEeqnarraynumspace \\
& = & - \frac{i}{8}\overline{\Psi} \big( 4\eta^{ab} \gamma^c - \gamma^c \gamma^a \gamma^b \big) \Big( \delta \big( \mathrm{d}\theta_a(v_b, v_c)  * \! 1 \big) \Big)  \Psi \label{eq197}
\end{IEEEeqnarray}
The first term on the right of \eqref{eq197} is, 
\begin{IEEEeqnarray}{rCl}
- \frac{i}{8}\overline{\Psi}4\eta^{ab} \gamma^c \Big(\delta \big( \mathrm{d}\theta_a(v_b, v_c) * \! 1 \big) \Big)  \Psi & = & \frac{i}{2}\overline{\Psi} \gamma^c  \Big( \delta \big( \mathrm{d}\theta^a(v_c, v_a) * \! 1 \big) \Big) \Psi \\
& = & \frac{i}{2}\overline{\Psi} \gamma^c ( \delta \mathrm{d} \! * \! \theta_c ) \Psi \\
\mathrm{d} \left( \frac{i}{2}\overline{\Psi}  \gamma^c (\delta \! * \!  \theta_c )\Psi \right) & = 
& \frac{i}{2} \mathrm{d}\overline{\Psi} \wedge \gamma^c(\delta \! * \!  \theta_c )\Psi + \frac{i}{2}\overline{\Psi} \gamma^c(\mathrm{d}\delta \! * \!  \theta_c )\Psi\:{+} \notag \\
& & {+}\:\frac{i}{2}\overline{\Psi}  \gamma^c \mathrm{d}\Psi \wedge \delta \! * \!  \theta_c 
\end{IEEEeqnarray}
so,
\begin{equation}
\frac{i}{2}\overline{\Psi} \gamma^c(\mathrm{d}\delta \! * \!  \theta_c )\Psi = - \frac{i}{2} \mathrm{d}\overline{\Psi} \wedge \gamma^c (\delta \! * \!  \theta_c ) \Psi - \frac{i}{2}\overline{\Psi}  \gamma^c \mathrm{d}\Psi \wedge \delta \! * \!  \theta_c  + \mathrm{d}(\ldots) \label{eq217}
\end{equation}
Combining \eqref{eq206} with the right side of \eqref{eq217} gives,
\begin{multline}
\frac{i}{2}\overline{\Psi} \gamma^a \mathrm{d}\Psi \wedge \delta \! * \! \theta_a - \frac{i}{2} \mathrm{d}\overline{\Psi} \wedge \gamma^a (\delta \! * \!  \theta_a ) \Psi = \\
= \delta\theta_m \wedge \left( \frac{i}{2} \eta^{mn}\overline{\Psi} \gamma^a \mathrm{d}  \Psi(v_a) - \frac{i}{2}\overline{\Psi} \gamma^n \mathrm{d} \Psi(v^m) \:{-} \right. \\
{-}\:\frac{i}{2} \eta^{mn}\mathrm{d}\overline{\Psi}(v_a) \gamma^a \Psi +  \frac{i}{2}\mathrm{d}\overline{\Psi}(v^m) \gamma^n \Psi \bigg) * \theta_n \label{eq219}
\end{multline}
I have,
\begin{IEEEeqnarray}{rCl}
\IEEEeqnarraymulticol{3}{l}{
\Big(\frac{i}{2}\eta^{mn}\overline{\Psi} \gamma^a \mathrm{d}  \Psi(v_a) -  \frac{i}{2} \overline{\Psi} \gamma^n \mathrm{d}  \Psi(v^m) \Big)^{\dagger} = } \notag \\
\qquad \qquad \qquad & = & -\frac{i}{2}\eta^{mn} \mathrm{d} \Psi^{\dagger}(v_a) \gamma^{a\dagger} \gamma^{0\dagger} \Psi + \frac{i}{2}\mathrm{d} \Psi^{\dagger}(v^m) \gamma^{n\dagger} \gamma^{0\dagger} \Psi \\
& = & -\frac{i}{2}\eta^{mn}\mathrm{d}\Psi^{\dagger}(v_a)\gamma^{a\dagger} (-\gamma^{0}) \Psi +\frac{i}{2}\mathrm{d}  \Psi^{\dagger}(v^m) \gamma^{n\dagger} (-\gamma^{0}) \Psi \IEEEeqnarraynumspace \\
& = & -\frac{i}{2}\eta^{mn}\mathrm{d}\overline{\Psi}(v_a) \gamma^{a} \Psi +\frac{i}{2}\mathrm{d}  \overline{\Psi}(v^m)\gamma^{n} \Psi
\end{IEEEeqnarray}
so the right side of \eqref{eq219} becomes,
\begin{equation}
\delta\theta_m \wedge \mathrm{Re} \Big( i\eta^{mn}\overline{\Psi} \gamma^a \mathrm{d}  \Psi(v_a) -  i\overline{\Psi} \gamma^n \mathrm{d} \Psi(v^m) \Big) * \! \theta_n
\end{equation}
The second term on the right of \eqref{eq197} is,
\begin{IEEEeqnarray}{rCl}
\IEEEeqnarraymulticol{3}{l}{ \frac{i}{8}\overline{\Psi}\gamma^c \gamma^a \gamma^b \big( \delta (\mathrm{d}\theta_a(v_b, v_c) * \! 1) \big) \Psi  =  \frac{i}{8}\overline{\Psi}\gamma^c \gamma^a \gamma^b \Psi \delta \big( \mathrm{d}\theta_a \wedge *(\theta_b \wedge \theta_c) \big) } \\ 
\qquad \qquad \qquad & = & \frac{i}{8}\overline{\Psi}\gamma^c \gamma^a \gamma^b \Psi  \Big( \mathrm{d} \delta \theta_a \wedge *(\theta_b \wedge \theta_c) + \mathrm{d} \theta_a \wedge \delta \! * \!  (\theta_b \wedge \theta_c) \Big) \IEEEeqnarraynumspace \label{eq198}
\end{IEEEeqnarray}
I have,
\begin{multline}
\mathrm{d} \Big( \overline{\Psi}\gamma^c \gamma^a \gamma^b \Psi  \big( \delta \theta_a \wedge *(\theta_b \wedge \theta_c) \big) \Big) = (\mathrm{d} \overline{\Psi}) \gamma^c \gamma^a \gamma^b \Psi \wedge \delta \theta_a \wedge *(\theta_b \wedge \theta_c)\:{+} \\
{+}\:\overline{\Psi}\gamma^c \gamma^a \gamma^b \mathrm{d}\Psi \wedge \delta \theta_a \wedge *(\theta_b \wedge \theta_c)  + \overline{\Psi}\gamma^c \gamma^a \gamma^b \Psi \, \mathrm{d} \delta \theta_a \wedge *(\theta_b \wedge \theta_c) \:{-}  \\ 
{-}\:\overline{\Psi}\gamma^c \gamma^a \gamma^b \Psi  \delta \theta_a \wedge \mathrm{d} \! * \! (\theta_b \wedge \theta_c)
\end{multline}
so,
\begin{multline}
\overline{\Psi}\gamma^c \gamma^a \gamma^b \Psi \, \mathrm{d} \delta \theta_a \wedge *(\theta_b \wedge \theta_c) = 
\delta \theta_a \wedge \Big( (\mathrm{d}\overline{\Psi})\gamma^c \gamma^a \gamma^b \Psi \wedge *(\theta_b \wedge \theta_c)\:{+}  \\ 
{+}\:\overline{\Psi}\gamma^c \gamma^a \gamma^b \mathrm{d}\Psi \wedge *(\theta_b \wedge \theta_c)  + \overline{\Psi}\gamma^c \gamma^a \gamma^b \Psi \, \mathrm{d} \! * \! (\theta_b \wedge \theta_c)\Big)  + \mathrm{d}(\ldots) 
\end{multline}
The first term on the right of \eqref{eq198} becomes,
\begin{multline}
\delta \theta_a \wedge \frac{i}{8} \Big( (\mathrm{d}\overline{\Psi})\gamma^c \gamma^a \gamma^b \Psi \wedge *(\theta_b \wedge \theta_c) + \overline{\Psi}\gamma^c \gamma^a \gamma^b \mathrm{d}\Psi \wedge *(\theta_b \wedge \theta_c)\:{+} \\
{+}\:\overline{\Psi}\gamma^c \gamma^a \gamma^b \Psi \, \mathrm{d} \! * \! (\theta_b \wedge \theta_c)\Big) + \mathrm{d}(\ldots) \label{eq218}
\end{multline}
The first term of \eqref{eq218} is,
\begin{IEEEeqnarray}{rCl}
\IEEEeqnarraymulticol{3}{l}{\delta \theta_a \wedge \frac{i}{8} ( \mathrm{d} \overline{\Psi} ) \gamma^c \gamma^a \gamma^b \Psi \wedge *(\theta_b \wedge \theta_c) = \delta \theta_a \wedge \frac{i}{8} (\mathrm{d}\overline{\Psi}) \gamma^c \gamma^a \gamma^b \Psi \wedge \frac{\epsilon_{bcde}}{2} \theta^d \wedge \theta^e } \IEEEeqnarraynumspace \\
\qquad \qquad & = & \delta \theta_a \wedge \frac{i}{16} \mathrm{d}\overline{\Psi}(v_f) \gamma^c \gamma^a \gamma^b \Psi \epsilon_{bcde} \epsilon^{nfde} * \! \theta_n \\
& = & \delta \theta_m \wedge \frac{i}{8} \mathrm{d}\overline{\Psi}(v_f) \big( \gamma^f\gamma^m\gamma^n - \gamma^n\gamma^m\gamma^f \big) \Psi * \! \theta_n 
\end{IEEEeqnarray}
Similarly, \eqref{eq218} becomes,
\begin{multline}
\delta \theta_m \wedge \left( \frac{i}{8} \mathrm{d}\overline{\Psi}(v_f) \big( \gamma^f\gamma^m\gamma^n - \gamma^n\gamma^m\gamma^f \big) \Psi * \! \theta_n\:{+} \right.  \\
{+}\:\frac{i}{8} \overline{\Psi} \big( \gamma^f\gamma^m\gamma^n - \gamma^n\gamma^m\gamma^f \big) \mathrm{d}\Psi(v_f) * \! \theta_n + \frac{i}{8} \overline{\Psi}\gamma^c \gamma^m \gamma^b \Psi \, \mathrm{d} \! * \! (\theta_b \wedge \theta_c) \bigg) + \mathrm{d}(\ldots) \label{eq226}
\end{multline}
I have,
\begin{IEEEeqnarray}{rCl}
\IEEEeqnarraymulticol{3}{l}{
\left( \frac{i}{8} \mathrm{d}\overline{\Psi}(v_f) \big( \gamma^f\gamma^m\gamma^n - \gamma^n\gamma^m\gamma^f \big) \Psi \right)^{\dagger} = } \notag \\
\qquad \qquad \qquad & = & \frac{-i}{8} \Psi^{\dagger}\big( \gamma^{n\dagger} \gamma^{m\dagger} \gamma^{f\dagger} - \gamma^{f\dagger} \gamma^{m\dagger} \gamma^{n\dagger} \big) \gamma^{0\dagger}\mathrm{d}\Psi(v_f) \\
& = & \frac{-i}{8} \Psi^{\dagger}\big( \gamma^{n\dagger} \gamma^{m\dagger} \gamma^{f\dagger} - \gamma^{f\dagger} \gamma^{m\dagger} \gamma^{n\dagger} \big) (-\gamma^{0})\mathrm{d}\Psi(v_f) \IEEEeqnarraynumspace \\
& = & \frac{-i}{8}\Psi^{\dagger} \gamma^{0} \big( \gamma^{n} \gamma^{m} \gamma^{f} - \gamma^{f} \gamma^{m} \gamma^{n} \big) \mathrm{d}\Psi(v_f) \\
& = & \frac{i}{8}\overline{\Psi} \big( \gamma^{f} \gamma^{m} \gamma^{n} - \gamma^{n} \gamma^{m} \gamma^{f}\big) \mathrm{d}\Psi(v_f) 
\end{IEEEeqnarray}
so \eqref{eq226} becomes,
\begin{gather}
\delta \theta_m \wedge \left( \mathrm{Re} \left(\frac{i}{4} \mathrm{d}\overline{\Psi}(v_f) \big( \gamma^f\gamma^m\gamma^n - \gamma^n\gamma^m\gamma^f \big) \Psi \right)* \! \theta_n \: + \right. \notag \\
+ \: \frac{i}{8} \overline{\Psi}\gamma^c \gamma^m \gamma^b \Psi \, \mathrm{d} \! * \! (\theta_b \wedge \theta_c) \bigg)  + \mathrm{d}(\ldots)
\end{gather}
The second term on the right of \eqref{eq198} is,
\begin{IEEEeqnarray}{rCl}
\frac{i}{8}\overline{\Psi}\gamma^c \gamma^a \gamma^b \Psi \, \mathrm{d} \theta_a \wedge \delta \! * \!  (\theta_b \wedge \theta_c) & = &  \frac{i}{8}\overline{\Psi}\gamma^c \gamma^a \gamma^b \Psi \, \mathrm{d} \theta_a \wedge \delta \left( \frac{\epsilon_{bcde}}{2} \theta^d \wedge \theta^e \right) \\
& = & \frac{i}{8}\overline{\Psi}\gamma^c \gamma^a \gamma^b \Psi \, \mathrm{d} \theta_a \wedge \epsilon_{bcde} \delta \theta^d \wedge \theta^e \\
& = & \frac{i}{8}\overline{\Psi}\gamma^c \gamma^a \gamma^b \Psi \, \mathrm{d} \theta_a \wedge \epsilon_{bcde} \eta^{dm} \delta \theta_m \wedge \theta^e \\
& = & \delta \theta_m \wedge \frac{i}{8}\overline{\Psi}\gamma^c \gamma^a \gamma^b \Psi \, \mathrm{d} \theta_a \wedge \epsilon_{bcde} \eta^{dm} \theta^e \IEEEeqnarraynumspace
\end{IEEEeqnarray}

The last term of \eqref{eq187} is,
\begin{IEEEeqnarray}{rCl}
i \overline{\Psi}  m \Psi \delta \! * \!  1 & = & i m \overline{\Psi} \Psi \delta \theta^m \wedge * \theta_m \\
& = & \delta \theta_m \wedge i m \overline{\Psi}\Psi \eta^{mn} * \! \theta_n
\end{IEEEeqnarray}

The variation of the Lagrangean density can be expressed as,
\begin{equation}
\delta \mathfrak{L} = \delta \theta_m \wedge T^m + \mathrm{d}(\ldots)
\end{equation}
with $T^m$ being the stress-energy tensor. Gathering terms, it is,
\begin{multline}
T^m = \bigg(F^{ma} F^n_{\phantom{n}a} - \frac{1}{4} \eta^{mn} F^{ab}F_{ab} 
- e \big( \eta^{mn}\overline{\Psi}\gamma^a \Psi A(v_a) - \overline{\Psi} \gamma^n \Psi A(v^m) \big)\:{+} \\ 
 {+}\:\mathrm{Re} \big(i \eta^{mn}\overline{\Psi} \gamma^a \mathrm{d}  \Psi(v_a) - i \overline{\Psi} \gamma^n \mathrm{d}  \Psi(v^m)\big)\:{+} \\
{+}\:\mathrm{Re}\big(\frac{i}{4} \overline{\Psi} (\gamma^f\gamma^m\gamma^n - \gamma^n\gamma^m\gamma^f) \mathrm{d}\Psi(v_f)\big) + i m \eta^{mn} \overline{\Psi}\Psi \bigg) * \! \theta_n\:{+} \\
{+}\:\frac{i}{8}\overline{\Psi}\gamma^c \gamma^m \gamma^b \Psi \, \mathrm{d} \! * \! (\theta_b \wedge \theta_c) 
+ \frac{i}{8}\overline{\Psi}\gamma^c \gamma^a \gamma^b \Psi \, \mathrm{d} \theta_a \wedge \epsilon_{bcde} \eta^{dm} \theta^e
\end{multline}
$T$ can also be written as,
\begin{gather}
T^{mn} = - * \! (\theta^n \wedge T^m)
\end{gather}
In flat spacetime with the Minkowski metric and with the coframe being the coordinate differentials, $T$ becomes,
\begin{multline}
T^{mn} = F^{ma} F^n_{\phantom{n}a} - \frac{1}{4} \eta^{mn} F^{ab}F_{ab} +  \eta^{mn}j^a A_a - j^n A^m\:{+} \\
{+}\:\mathrm{Re}\big(i \eta^{mn}\overline{\Psi} \gamma^a \partial_a  \Psi - i \overline{\Psi} \gamma^n \partial^m \Psi \big)\:{+} \\
{+}\:\mathrm{Re} \left( \frac{i}{4} \overline{\Psi} (\gamma^f\gamma^m\gamma^n - \gamma^n\gamma^m\gamma^f) \partial_f\Psi \right) + i m \eta^{mn} \overline{\Psi} \Psi  
\end{multline}
$i\overline{\Psi} \Psi$ is real because $\gamma^0$ is a diagonal matrix. I use,
\begin{gather}
\overline{\Psi}\gamma^a \partial_a  \Psi - ij^a A_a = - m \overline{\Psi} \Psi
\end{gather}
and, 
\begin{IEEEeqnarray}{rCl}
\IEEEeqnarraymulticol{3}{l}{ \overline{\Psi} \big( \gamma^f\gamma^m\gamma^n - \gamma^n\gamma^m\gamma^f \big) \partial_f\Psi = \overline{\Psi} \big( 2 \eta^{fm} \gamma^n - 2 \eta^{fn} \gamma^m + [\gamma^m, \gamma^n] \gamma^f \big) \partial_f\Psi } \IEEEeqnarraynumspace \\
\IEEEeqnarraymulticol{3}{l}{ \mathrm{Re} \left( \frac{i}{4} \overline{\Psi} [\gamma^m, \gamma^n] \gamma^f \partial_f\Psi \right) = \mathrm{Re} \left( \frac{i}{4} \overline{\Psi} [\gamma^m, \gamma^n] \big( -ie \gamma^a A_a - m \big) \Psi \right) } \\
\qquad \qquad & = & \mathrm{Re} \left( \frac{1}{4} \overline{\Psi} \left( 2e\gamma^m A^n - 2e\gamma^n A^m + \frac{e}{3}[\gamma^m, \gamma^n , \gamma^a] A_a \:{-} \right. \right. \notag \\
& & {-}\:im [\gamma^m, \gamma^n] \Big) \Psi \bigg) \\
& = & \frac{1}{2} \big( - j^m A^n + j^n A^m \big)\:{+} \notag \\
& & {+}\:\mathrm{Re} \left( \overline{\Psi} \left(\frac{e}{12}[\gamma^m, \gamma^n , \gamma^a] A_a - \frac{i}{4} m [\gamma^m, \gamma^n]\right) \Psi \right)
\end{IEEEeqnarray}
$\gamma^0 [\gamma^m, \gamma^n]$ is Hermitean:
\begin{IEEEeqnarray}{rCl}
\big(\gamma^0 [\gamma^m, \gamma^n]\big)^{\dagger} & = & [\gamma^{n\dagger}, \gamma^{m\dagger}]\gamma^{0\dagger} \\
& = & [\gamma^{n\dagger}, \gamma^{m\dagger}](-\gamma^0) \\
& = & -\gamma^0 [\gamma^n, \gamma^m] \\
& = & \gamma^0 [\gamma^m, \gamma^n]
\end{IEEEeqnarray}
so $\overline{\Psi}[\gamma^m, \gamma^n] \Psi$ is real. $\gamma^0[\gamma^m, \gamma^n , \gamma^a]$ is skew-Hermitean:
\begin{IEEEeqnarray}{rCl}
\big(\gamma^0[\gamma^m, \gamma^n , \gamma^a]\big)^{\dagger} & = & [\gamma^{a\dagger}, \gamma^{n\dagger}, \gamma^{m\dagger}]\gamma^{0\dagger} \\
& = & [\gamma^{a\dagger}, \gamma^{n\dagger}, \gamma^{m\dagger}] (-\gamma^0) \\
& = & \gamma^0[\gamma^a, \gamma^n, \gamma^m] \\
& = & -\gamma^0 [\gamma^m, \gamma^n , \gamma^a]
\end{IEEEeqnarray}
so $\overline{\Psi} [\gamma^m, \gamma^n , \gamma^a] \Psi$ is imaginary or $0$. $T$ becomes,
\begin{multline}
T^{mn} = F^{ma} F^n_{\phantom{n}a} - \frac{1}{4} \eta^{mn} F^{ab}F_{ab}
- \frac{1}{2} \big( j^m A^n + j^n A^m \big)\:{+} \\
{+}\:\mathrm{Re} \left( \frac{-i}{2}\overline{\Psi} \big( \gamma^m \partial^n + \gamma^n \partial^m \big) \Psi \right)
\end{multline}
So the energy density is,
\begin{equation}
T^{00} = \frac{1}{2}\big(\mathbf{E}^2 + \mathbf{B}^2\big) + j^0A_0 + \mathrm{Re} \left(-i\Psi^{\dagger} \partial_0 \Psi \right)
\end{equation}

\section*{Appendix} \label{Appendix}

$\omega$ has the same values as the rotational part of the torsion-free Cartan connection $(\theta, \omega): \mathrm{T}(\mathcal{M}^4) \rightarrow \mathfrak{euc}(3,1)$ because $\omega$ is antisymmetric and follows the equation, 
\begin{gather}
\mathrm{d}\theta^a - \omega^a_{\phantom{a}b} \wedge \theta^b = 0 \label{eq26}
\end{gather}
To prove \eqref{eq26}:
\begin{IEEEeqnarray}{rCl}
\IEEEeqnarraymulticol{3}{l}{
\mathrm{d}\theta^a(v_c, v_d) - (\omega^a_{\phantom{a}b} \wedge \theta^b)(v_c, v_d) = } \notag \\
\qquad \qquad & = & \mathrm{d}\theta^a(v_c, v_d) - \omega^a_{\phantom{a}b}(v_c) \theta^b(v_d) + \omega^a_{\phantom{a}b}(v_d) \theta^b(v_c) \\
& = & \mathrm{d}\theta^a(v_c, v_d) - \omega^a_{\phantom{a}d}(v_c) + \omega^a_{\phantom{a}c}(v_d) \\
& = & \mathrm{d}\theta^a(v_c, v_d) - \frac{1}{2} \Big( \mathrm{d}\theta_c(v^a, v_d) + \mathrm{d}\theta^a(v_c, v_d) - \mathrm{d}\theta_d(v_c, v^a) \Big)\:{+} \notag \\
& & {+}\:\frac{1}{2} \Big( \mathrm{d}\theta_d(v^a, v_c) + \mathrm{d}\theta^a(v_d, v_c) - \mathrm{d}\theta_c(v_d, v^a) \Big) \\
& = & \mathrm{d}\theta^a(v_c, v_d) - \mathrm{d}\theta^a(v_c, v_d) \\
& = & 0
\end{IEEEeqnarray}
To prove that \eqref{eq26} is the formula for the torsion: The formula for the curvature of a Cartan connection $\xi$ is \cite{Sharpe},
\begin{gather}
\mathrm{d} \xi + \frac{1}{2}[\xi, \xi]
\end{gather}
with,
\begin{gather}
[\xi, \chi](U, V) = [\xi(U), \chi(V)] - [\xi(V), \chi(U)]
\end{gather}
with the commutator evaluated in a Lie algebra. For the Cartan connection $(\theta, \omega): \mathrm{T}(\mathcal{M}^4) \rightarrow \mathfrak{euc}(3,1)$, the torsion is the translational part of the curvature:
\begin{gather}
\mathrm{d} \theta + \frac{1}{2}[\theta, \omega] + \frac{1}{2}[\omega,\theta] 
\end{gather}
To simplify:
\begin{IEEEeqnarray}{rCl}
\IEEEeqnarraymulticol{3}{l}{
(\mathrm{d} \theta + \frac{1}{2}[\theta, \omega] + \frac{1}{2}[\omega,\theta])(U, V) = } \\
\qquad \qquad & = & \mathrm{d} \theta(U, V) + \frac{1}{2}[\theta(U), \omega(V)] - \frac{1}{2}[\theta(V), \omega(U)]\:{+} \notag \\
& & {+}\:\frac{1}{2}[\omega(U),\theta(V)] - \frac{1}{2}[\omega(V),\theta(U)] \\
& = & \mathrm{d} \theta(U, V) + [\omega(U), \theta(V)] - [\omega(V), \theta(U)] \IEEEeqnarraynumspace \\
& = & (\mathrm{d} \theta + [\omega, \theta])(U, V)
\end{IEEEeqnarray}
so the torsion is,
\begin{gather}
\mathrm{d} \theta + [\omega, \theta] \label{eq40}
\end{gather}
Some of the generators of $\mathrm{Euc}(3, 1)$ are,
\begin{gather}
\begin{array}{cc}
J^{12} = \left[\begin{array}{ccccc}
0 & 0 & 0 & 0 & 0 \\
0 & 0 & -1 & 0 & 0 \\
0 & 1 & 0 & 0 & 0 \\
0 & 0 & 0 & 0 & 0 \\
0 & 0 & 0 & 0 & 0 \end{array} \right]
&
t^2 = \left[\begin{array}{ccccc}
0 & 0 & 0 & 0 & 0 \\
0 & 0 & 0 & 0 & 0 \\
0 & 0 & 0 & 0 & 1 \\
0 & 0 & 0 & 0 & 0 \\
0 & 0 & 0 & 0 & 0 \end{array} \right]
\end{array}
\end{gather}
They follow the commutation relations,
\begin{gather}
[J^{ab}, t^c] = -\eta^{bc} t^a + \eta^{ac} t^b 
\end{gather}
In this appendix, the values of $\omega$ are combinations of $J$ matrices:
\begin{gather}
\omega = \frac{1}{2}J^{ab} \omega_{ab}
\end{gather}
The rotation generators are duplicated (for example, there is $J^{21}$ in addition to $J^{12}$), hence the factor $\frac{1}{2}$. Formula \eqref{eq40} applied to vectors $U$ and $V$ gives,
\begin{IEEEeqnarray}{rCl}
(\mathrm{d} \theta + [\omega, \theta])^a (U, V) & = & \mathrm{d} \theta^a(U, V) + [\omega(U), \theta(V)]^a - [\omega(V), \theta(U)]^a \IEEEeqnarraynumspace \\
& = & (\mathrm{d} \theta^a - \frac{1}{2} \omega^a_{\phantom{a}b} \wedge \theta^b + \frac{1}{2} \omega_b^{\phantom{b}a} \wedge \theta^b)(U, V) \\
& = & (\mathrm{d} \theta^a - \omega^a_{\phantom{a}b} \wedge \theta^b)(U, V)
\end{IEEEeqnarray}
So the torsion is,
\begin{gather}
\mathrm{d} \theta^a - \omega^a_{\phantom{a}b} \wedge \theta^b
\end{gather}

\pagebreak


\begin{thebibliography}{9}

\bibitem{Visinescu} I. I. Cot\u{a}escu, M. Visinescu. ``Symmetries and supersymmetries of the Dirac operators in curved spacetimes''. \emph{Progress in General Relativity and Quantum Cosmology Research}, Nova Science, N.Y. (2007), pp.109-166. \href{http://arxiv.org/abs/hep-th/0411016}{arXiv:hep-th/0411016}
\bibitem{Weyl} Hermann Weyl. ``A Remark on the Coupling of Gravitation and Electron''. \emph{Phys. Rev. 77}, 699 (1950). 
\bibitem{Marchuk} N. G. Marchuk. ``A gauge model with spinor group for a description of local interaction of a fermion with electromagnetic and gravitational fields''. \emph{Nuovo Cimento}, 115B, 11, 2000.  \href{http://arxiv.org/abs/math-ph/9912004}{arXiv:math-ph/9912004}
\bibitem{Marchuk2} N. G. Marchuk. ``A coordinateless form of the Dirac equation''. \href{http://arxiv.org/abs/math-ph/0307042}{arXiv:math-ph/0307042}
\bibitem{Sharpe} R. W. Sharpe. \emph{Differential Geometry: Cartan's Generalization of Klein's Erlangen Program}. Springer.
\bibitem{Itin} Yakov Itin. ``Energy-momentum current for coframe gravity''. \emph{Class. Quant. Grav.} 19 (2002) 173-190. \href{http://arxiv.org/abs/gr-qc/0111036}{arXiv:gr-qc/0111036}
\end{thebibliography}
\end{document}